# Highly precise AMCW time-of-flight scanning sensor based on digital-parallel demodulation

Sung-Hyun Lee, Wook-Hyeon Kwon, Yoon-Seop Lim, Yong-Hwa Park
Department of Mechanical Engineering, Korea Advanced Institute of Science and Technology,
Daejeon 34141, Korea

## ABSTRACT

In this paper, a novel amplitude-modulated continuous wave (AMCW) time-of-flight (ToF) scanning sensor based on digital-parallel demodulation is proposed and demonstrated in the aspect of distance measurement precision. Since digital-parallel demodulation utilizes a high-amplitude demodulation signal with zero-offset, the proposed sensor platform can maintain extremely high demodulation contrast. Meanwhile, as all cross correlated samples are calculated in parallel and in extremely short integration time, the proposed sensor platform can utilize a 2D laser scanning structure with a single photo detector, maintaining a moderate frame rate. This optical structure can increase the received optical SNR and remove the crosstalk of image pixel array. Based on these measurement properties, the proposed AMCW ToF scanning sensor shows highly precise 3D depth measurement performance. In this study, this precise measurement performance is explained in detail. Additionally, the actual measurement performance of the proposed sensor platform is experimentally validated under various conditions.

## 1. INTRODUCTION

3D depth information has been widely used to improve the performance of object recognition, which is one of key issues for the realization of fully automated intelligent systems [1–6]. For example, 3D depth images can be utilized in the object classification process of bin picking robots to improve the accuracy of sensing the position and orientation of objects [3]. Autonomous robots and vehicles utilize 3D depth images to improve the accuracy of mapping, localization, and object segmentation algorithms [1,2,4]. Meanwhile, 3D depth images of measured objects can be used for medical electronics to improve the interface familiarity between users and devices [5]. Even for drones flying around a dense downtown area, global mapping and sensing of surrounding obstacles can be greatly enhanced by utilizing 3D depth information [6]. Likewise, 3D depth information can be widely used to achieve the highly accurate object sensing technology of intelligent systems which are essential for solving automation issues in a wide variety of fields.

Among various 3D depth measurement technologies, light detection and ranging (LiDAR) solutions are widely used these days [7–14]. In general, LiDAR systems consist of an illumination source, scanner, optical lens, and photo detector(s) to measure the time-of-flight (ToF) of light signal [7–12]. When the light signal emitted from the illumination source is reflected from the object to the photo detector, a time delay between the emitted light and received light essentially occurs. This time delay is called ToF, which is measured directly by the time-to-digital converter (TDC) or indirectly estimated [12,14]. Direct pulse-based ToF measurement sensors, which are widely used for relatively long range measurement, scan the emitted light signal of a pulse waveform to an object scene using a fast scanner [7–12]. The ToF of the light signal is then directly measured using the TDC circuit whose measurement resolution should be up to picoseconds [12]. Although such a highly precise TDC circuit is possible due to the development of circuit fabrication, the cost is still an issue for direct pulse-based ToF sensors [12].

Alternatively, indirect ToF measurement sensors are also widely used for relatively short range up to 10 m [13-24]. Many indirect ToF measurement sensors utilize an amplitude-modulated continuous wave (AMCW) ToF measurement method. Once the light source is modulated at a specific modulation frequency of usually tens of MHz and illuminated over the entire object scene by diffuser lens, the modulated light signal is then reflected from the object. This reflected light signal is focused by a focusing lens onto the CMOS demodulation pixel array to be demodulated by each pixel [12,14,15,22,23]. At each demodulation pixel, usually four demodulation processes occur at 4 different photogates in the demodulation pixel with the phase shifted by 0°, 90°, 180°, 270° for each. These 4 demodulated results, known as cross correlated samples, are read and sampled by the readout circuit right after the demodulation process. These cross correlated samples are used to calculate the phase difference between the emitted light and received light, which can be converted into the measured distance. Likewise, many ToF cameras utilize such AMCW method and focal plane array structure which consists of CMOS demodulation pixels [12,14,15,22,23]. These ToF cameras show relatively precise distance measurement performance according to previous research [16,19,20]. Additionally, due to the progress of the CMOS image pixel fabrication process, the price per one image pixel has decreased, making ToF cameras cheaper than direct ToF LiDAR [12,14,17].

However, the measurement precision of conventional ToF cameras has limitations due to the limited demodulation contrast of the demodulation pixel [15,18,22,25]. When photoelectrons are generated by the received light signal in the semiconductor, some of these generated photoelectrons are transported in wrong direction [25]. Such loss of generated photoelectrons results in limited demodulation contrast, which is one of the main causes of distance measurement precision decrease in ToF cameras [15,18,22,25]. Except for demodulation contrast, the small fill factor of demodulation pixels is also problem which decreases the amount of received light [25]. Additionally, neighboring pixels can interfere with each other, causing an increase in unwanted noise [14,26]. All these limitations related to the demodulation contrast, fill factor, and crosstalk are inevitable in that the demodulation process of conventional ToF cameras depends entirely on the CMOS demodulation pixel array structure.

Meanwhile, the relatively low reflected optical illumination power per one image pixel of ToF cameras also affects the measurement precision. As conventional ToF cameras use a diffuser lens to scatter an illuminated light signal over the entire object scene, the reflected illumination power per one image pixel is relatively low [27]. To increase the received optical SNR per one demodulation pixel, the power of the illumination source should be high enough up to 1000 mW, which results in a large consumption of electric energy [22,23,26–31]. Moreover, even if there is enough illumination power, the received optical SNR per one pixel decreases if the image resolution is increased for the same intensity of illumination power. One alternative to improve the received optical SNR per one image pixel of ToF cameras is utilizing a 2D scanning structure with a single photo detector which focuses the total reflected light energy on the single active area of the photo detector. However, this scanning method is not compatible with conventional ToF cameras due the relatively long integration time per one demodulation pixel up to tens of milliseconds, which results in an extremely low frame rate for scanning over all demodulation pixels [19,27]. Although there exist some types of AMCW ToF scanning sensors, these sensors can be used in limited situations since these sensors can scan the object in only one direction [32,33].

To solve the problems related to the CMOS demodulation pixel array and low received optical SNR of conventional ToF cameras, a novel AMCW ToF scanning sensor platform with a single photo detector, which has been improved from previous works, is proposed and demonstrated in this paper [34]. Compared to the previous work, the measurement properties of the proposed sensor platform are theoretically analyzed in detail in the aspect of demodulation contrast and measurement precision [27,34,35]. Since the demodulation signal of the proposed sensor platform, which utilizes digital-parallel demodulation, has a sinusoidal waveform, the amplitude of cross correlation is not attenuated compared to other ToF cameras of which the demodulation signal has a square waveform [22,23,27,34,36,37]. Additionally, since the demodulation signal of the proposed sensor platform has zero-offset, the offset of cross correlation is attenuated to almost zero, which results in an extremely high demodulation contrast of the proposed sensor platform. Consequently, highly precise distance measurement is feasible using the proposed sensor platform even if the integration time is extremely short. Since all cross correlations based on the digitized signals are calculated in parallel by a real time processor, the acquisition time for one image frame is equal to the product of image resolution (number of pixels) and the integration time of a single pixel for the proposed sensor platform. According to this property, if the integration time is less than one microsecond, the time to acquire one image frame is less than 0.08 sec with QVGA image resolution, which is a moderate speed compared to other conventional ToF cameras [12,19]. Therefore, a 2D scanning structure with a single avalanche photodiode (APD) is adopted for the proposed sensor platform to achieve a high received optical SNR and the elimination of crosstalk with a moderate frame rate. The increased optical SNR and removal of crosstalk also contribute to the improved measurement precision of the proposed sensor platform compared to that of other conventional ToF cameras. In summary, adopting the digital-parallel demodulation method and a 2D scanning structure with single APD together, the proposed sensor platform shows high measurement precision and high quality of a 3D depth map with a moderate frame rate. To validate the proposed sensor platform's highly precise measurement performance, experiments with various conditions were conducted and are shown in this paper.

This paper is organized as follows: Section 2 describes the general operating principle of conventional ToF cameras. Section 3 describes the working principle and theoretical measurement properties of the AMCW scanning sensor platform proposed in this paper. In Section 4, distance measurement performance and 3D images are demonstrated in various experimental conditions. Section 5 presents the conclusion of this paper.

## 2. OPERATING PRINCIPLE OF TIME-OF-FLIGHT CAMERAS

The main principle of a general AMCW ToF camera is to measure the time delay between the illuminated light signal and received light signal [12,14,18]. This ToF measurement process requires an illumination source, optical lens, optical detector, and control electronics, as shown in Fig. 1.

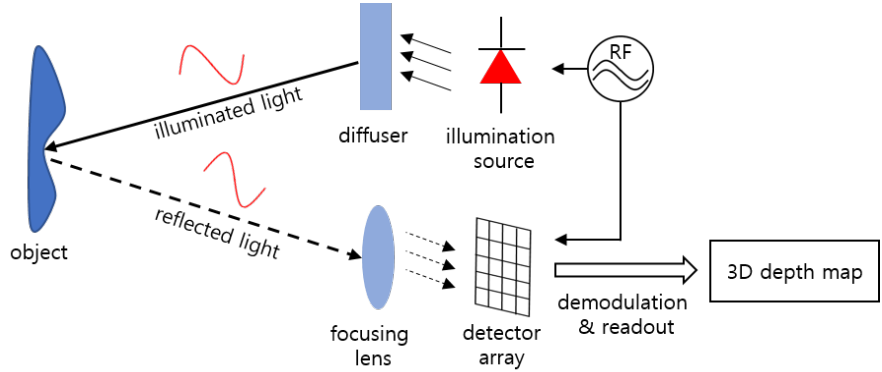

**Fig. 1.** General operating principle of existing ToF measurement cameras

Generally, AMCW ToF cameras utilize a phase correlation method to estimate the phase difference between the illuminated light signal and received light signal, which can be converted into the distance of an object. Specifically, the received light signal is demodulated by CMOS demodulation pixels to acquire several (typically four) cross correlated samples. All related mathematical expressions are as follows [12,18,27,35]:

$$C(\varphi_n) = \lim_{T_{int} \to \infty} \frac{1}{T_{int}} \int_{-T_{int}/2}^{T_{int}/2} r(t) \cdot s(t + \frac{\varphi_n}{2\pi f}) dt, \quad n = 0, ..., 3 \tag{1}$$

$$A = \frac{\sqrt{\{C(\varphi_0) - C(\varphi_2)\}^2 + \{C(\varphi_3) - C(\varphi_1)\}^2}}{2} \tag{2}$$

$$B = \frac{C(\varphi_0) + C(\varphi_1) + C(\varphi_2) + C(\varphi_3)}{4} \tag{3}$$

$$\varphi = \arctan\left(\frac{C(\varphi_3) - C(\varphi_1)}{C(\varphi_0) - C(\varphi_2)}\right) \tag{4}$$

where $C(\varphi_n)$ is the cross correlation of signals, $T_{int}$ is the integration time, $r(t)$ is the received light signal, $s(t)$ is the demodulation signal, $f$ is the modulation frequency, $\varphi_n$ is the phase shift of the demodulation signal which is chosen as $\varphi_0 = 0, \varphi_1 = \pi/2, \varphi_2 = \pi, \varphi_3 = 3\pi/2$ for each cross correlation, $A$ is the amplitude of the cross correlation, $B$ is the offset of the cross correlation, and $\varphi$ is the phase of the cross correlation. Using the phase in Eq. (4), distance is calculated as follows:

$$d = \frac{1}{2} \cdot \left(\frac{c \cdot \varphi}{2\pi f}\right) \tag{5}$$

where $c$ is the velocity of light and $f$ is the modulation frequency. This AMCW method is widely used for various ToF cameras, such as Kinect V2, SR series, and PMD series [16,19,24]. Due to the progress of CMOS sensor fabrication, a relatively high measurement precision and cheap price have become feasible for ToF cameras [12,14,17]. However, there are still limitations related to the demodulation contrast and fill factor of pixels in ToF cameras, which results in reduced measurement precision [15,18,22,25]. Additionally, as many types of ToF cameras utilize a focal plane array structure which consists of a number of demodulation pixels, unwanted crosstalk from each pixel inevitably occurs, which results in a fixed pattern noise on a 3D depth image [14,26]. Meanwhile, to illuminate light over the entire object scene using a diffuser, an approximately 1000 mW illumination source should be adopted, leading to high electric energy consumption and low received optical SNR per one demodulation pixel [22,23,26–31]. To eliminate these measurement limitations, the overall optical structure of the platform should be modified.

This paper suggests a scanning type AMCW ToF sensor platform which utilizes a four tap digital-parallel demodulation in contrast to the previous sequential demodulation scheme. A detailed description of this proposed sensor platform is presented in next section.

## 3. AMCW TOF SCANNING SENSOR PLATFORM BASED ON DIGITAL-PARALLEL DEMODULATION

### 3.1 Working principle of AMCW ToF scanning sensor platform based on digital-parallel demodulation

A novel 2D scanning type AMCW ToF sensor platform with a single photodetector is developed using a digital-parallel demodulation process to eliminate the limitations of conventional ToF cameras described in Section 2. The block diagram of the proposed AMCW ToF sensor platform is presented in Fig. 2.

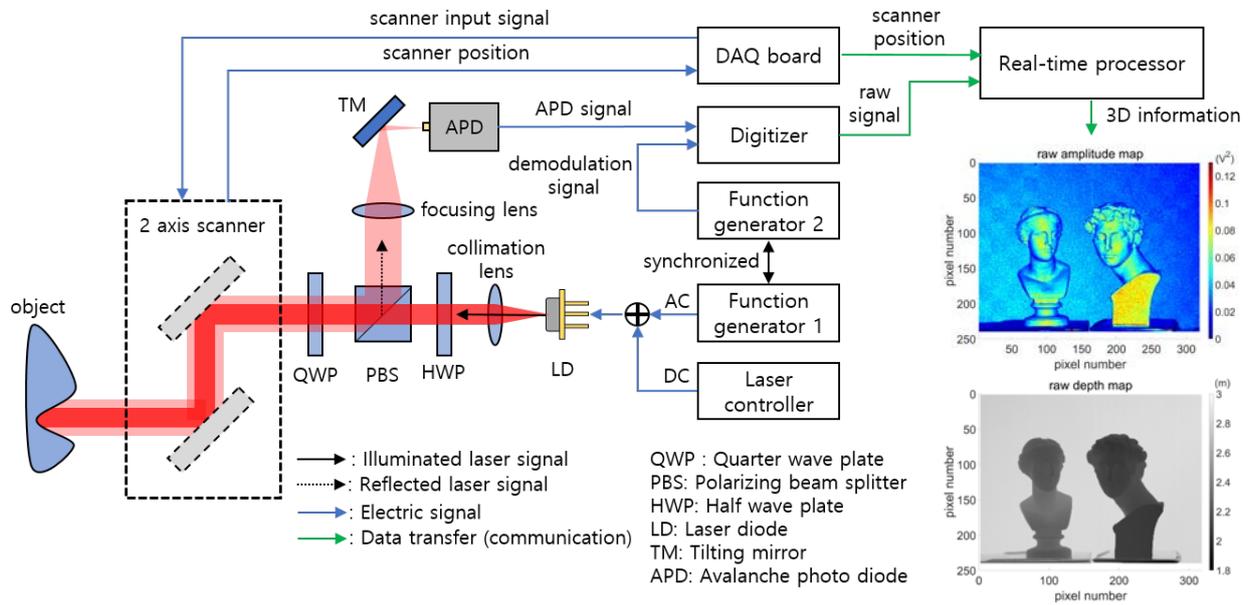

**Fig. 2.** Block diagram of AMCW ToF scanning sensor platform based on digital-parallel demodulation

Two function generators (NI-PXI-5404, 5406) are synchronized by an internal clock embedded in the chassis (NI-PXIe-1082) at a modulation frequency of 31.25 MHz to generate sinusoidal electric signals. One of the sinusoidal electric signals is transmitted to the laser diode module (Thorlabs, TCLDM9) to modulate the laser diode with a wavelength of 852 nm (Thorlabs, M9-852-0100). This sinusoidal electric signal is combined with the DC offset electric signal from the laser controller (Arroyo instruments, ComboSource 6340) in the bias T circuit which is inside of the laser diode module. This combined sinusoidal electric signal with DC offset is directly transmitted to the laser diode. From the laser diode, a modulated laser signal is directly illuminated to the collimation lens (Thorlabs, A240TM-B). This collimated laser signal is p-polarized by the HWP (Thorlabs, WPH10M-850) to pass through the PBS. After passing through the PBS (Thorlabs, CCM1-PBS252), the polarization status of the laser signal is converted into circular status by the QWP (Thorlabs, WPQ20ME-850-SP). This laser signal is scanned on the object in horizontal and vertical directions by a two-axis scanner (Thorlabs, GVS 012). The scanned laser signal is then reflected back to the sensor platform following the same optical path to reach PBS. Parts of this reflected laser signal are directly focused on the active area of APD (Thorlabs, APD120A) by the focusing lens (Thorlabs, LA4725-B) and tilting mirror (Thorlabs, BB1-E03). Meanwhile, a fast digitizer (NI-PXIe-5160) digitizes the laser signal received by APD and demodulation signal from the other function generator using two channels at a sampling frequency of 625 MHz. These digitized signals are transmitted to the real-time processor (NI-PXIe-8880) by internal communication of the chassis to be used for the calculation of four cross correlated samples using Eq. (1). When four cross correlated samples are calculated, all these calculations are processed in parallel by real time-LabVIEW software. All the processes related to signal digitization and calculation of cross correlated samples are referred to as the *digital-parallel demodulation process* in this work. To generate a 3D depth map, scanning angles of a two-axis scanner are also transmitted to a real time processor after being digitized by the DAQ module (NI-PXIe-7868R). At the final step, a real time processor constructs an amplitude map and 3D depth map, and this 3D information is displayed on the host PC.

The 2D scanning structure with a single APD, shown in Fig. 2, has some structural advantages compared to the focal plane array structure as follows. Since a 2D scanning structure focuses the entire illuminated light signal on the specific point of object, the power of the reflected laser signal per one image pixel is increased. Additionally, since the reflected laser signal is completely focused on the active area of a single APD, there is very little optical energy loss of the reflected laser signal, which is not feasible for other ToF cameras due to fill factor limits. Based on such focusing properties, the 2D scanning structure with a single APD resultantly improves the received optical SNR per image pixel compared to that of the focal plane array structure with the same optical illumination power [7–10]. Additionally, as there is just one photodetector, there is no neighbor pixels in the system. This means that there is no crosstalk by neighbor pixels, which can suppress the fixed pattern noise frequently shown in other ToF cameras [14,26]. However, this 2D scanning structure is not compatible with conventional ToF cameras due to the relatively long integration time per one demodulation pixel up to tens of milliseconds, which results in an extremely low frame rate for scanning over all demodulation pixels. [19,27]. On the other hand, the AMCW ToF sensor platform proposed in this paper is compatible with 2D scanning structure due to the extremely short integration time of the digital-parallel demodulation method. The overall description of the digital-parallel demodulation method is shown in Fig. 3.

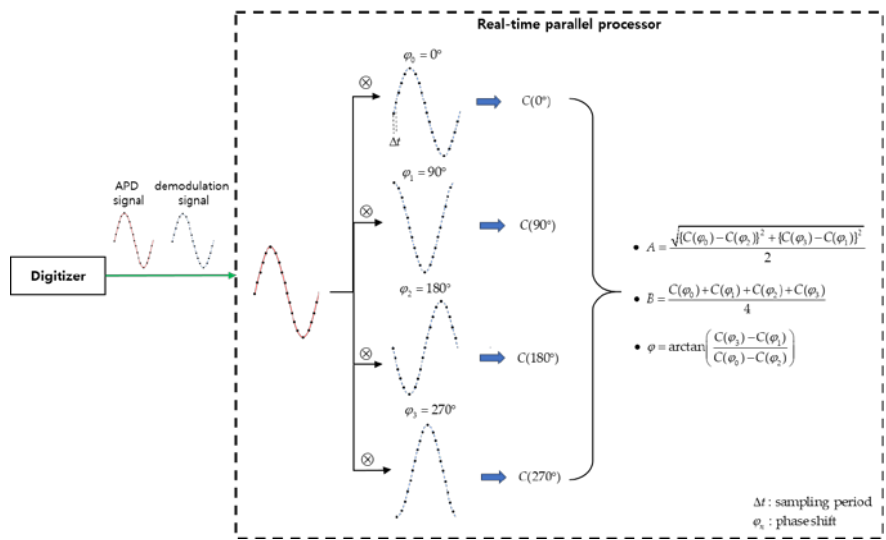

**Fig. 3.** Schematic of digital-parallel demodulation process using an APD signal and demodulation signal to acquire the amplitude, offset, and phase of cross correlated samples

According to Fig. 3, the digitizer transmits each APD signal and demodulation signal to the real time processor with sampling period $\Delta t$. The demodulation signal is then duplicated four times, being phase shifted by 0°, 90°, 180°, 270° for each in the real time processor. After four demodulation signals are generated in the digital domain, all cross correlations are calculated in parallel and in real time by the processor, as shown in Fig. 3. Using these 4 cross correlations, the phase is finally calculated using Eq. (4). This digital-parallel demodulation process is compatible with 2D scanning structure in that integration time can be radically decreased by controlling the length of the digitized signal to result in a moderate frame rate. Additionally, since all four cross correlations are calculated in parallel and in real time, only one integration time is enough for one distance measurement. Consequently, time to acquire one image frame of the proposed AMCW ToF scanning sensor platform is almost same as the product of image resolution (number of pixels) and integration time per pixel. Specifically, if the integration time is set to 800 nsec and image resolution is in QVGA for each, the time to acquire one image frame is approximately 0.0614 sec, which is a moderate speed compared to that of other ToF cameras [19,27]. Based on this digital-parallel demodulation process, the sensor platform proposed in this paper utilizes a 2D scanning structure with a single APD, as shown in Fig. 2, to take advantage of the improved measurement precision related to the optical structure.

Meanwhile, except for the structural advantages of the 2D scanning structure with single APD, AMCW ToF scanning sensor platform in Fig. 2 has another advantage related to the demodulation contrast. This property is explained in detail in Section 3.2.

### 3.2 Analysis of digital-parallel demodulation method using Fast Fourier Transform

As described in the previous section, the digital-parallel demodulation process utilizes the digitization of the received laser signal and demodulation signal to calculate four cross correlated samples in parallel using a real time processor. This correlation in the digital domain has strong advantages because demodulation contrast is extremely high. To analyze the demodulation contrast of the proposed sensor platform, signal analysis was conducted, as described in Fig. 4.

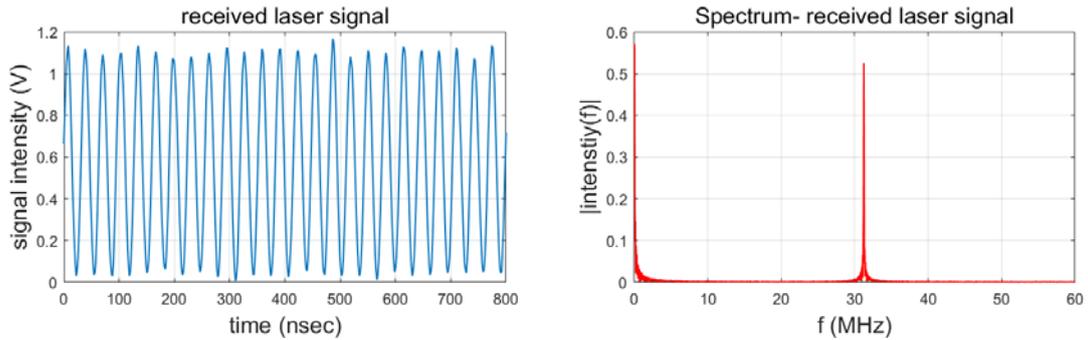

(a) Received laser signal

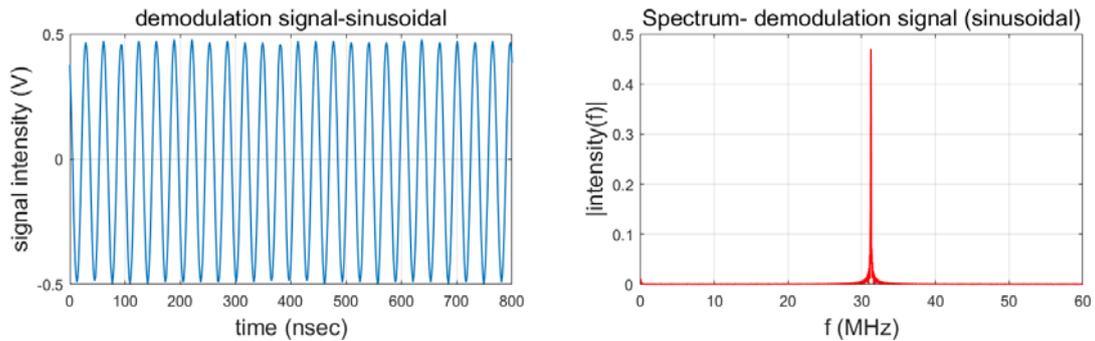

(b) Sinusoidal demodulation signal with zero offset

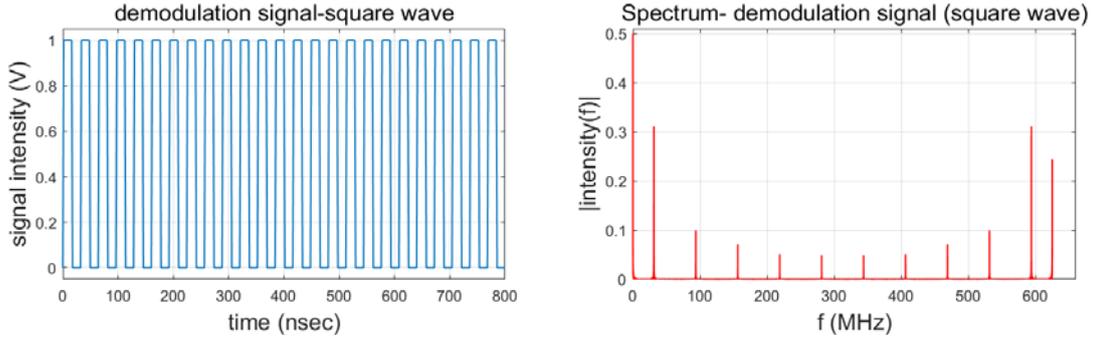

(c) Square wave demodulation signal with 0.5 offset

**Fig. 4.** Plots of signals and FFT spectrum for each signal. (a): plot of received laser signal, (b): plot of sinusoidal demodulation signal, (c): plot of square wave demodulation signal

The experimental conditions of Fig. 4 are as follows: the record length for each signal is 10,000, the number of data points for Fast Fourier Transform (FFT) is 16,384, the modulation frequency is 31.25 MHz, the peak-to-peak value of the sinusoidal demodulation signal is close to 1, the sampling rate is 625 MHz, the measured distance is about 1.3 m, and the measured object is white paper. The received laser signal in Fig. 4(a) is the actual received laser signal digitized as shown in Fig. 2. The spectrum of the received laser signal shows that the intensity of the spectrum at 0 MHz and 31.25 MHz is 0.572 and 0.526, respectively. The sinusoidal demodulation signal shown in Fig. 4(b) is also an actual digitized demodulation signal which is generated by the function generator shown in Fig. 2. The corresponding FFT spectrum of the sinusoidal demodulation signal shows that there is an extremely low offset of approximately 0.0099 at 0 MHz and a peak value of about 0.4704 at 31.25 MHz. Using these digitized signals based on the process shown in Fig. 3, the offset of cross correlation is attenuated to about 0.0057 according to the convolution theorem [38]. This indicates that the demodulation contrast for this case is extremely high compared to that of conventional ToF cameras, of which demodulation contrast is generally up to 0.60 [22,25,27]. The demodulation contrast is defined as follows: [15,25,27].

$$C_{demodulation} = \frac{A}{B} \quad (6)$$

where $C_{demodulation}$ is the demodulation contrast and $A$ and $B$ are defined in Eqs. (2) and (3). According to Eq. (6), the demodulation contrast calculated using the received laser signal and sinusoidal demodulation signal in Fig. 4 is about 43.2203. In conventional ToF cameras, which use CMOS demodulation pixels, the demodulation contrast cannot be higher than 1 due to the offset terms of the square wave demodulation signal. However, by attenuating the DC offset term of the demodulation signal to almost zero, it is possible to increase the demodulation contrast extremely high, up to 43.2203 in this experiment, using the digital-parallel demodulation method with a sinusoidal demodulation signal, as shown in Fig. 4(b). Since the measurement precision is proportional to the demodulation contrast, this digital-parallel demodulation method directly improves the measurement precision, which is validated by the experimental results shown in Section 4.

Meanwhile, the other demodulation signal shown in Fig. 4(c) is the ideal square waveform which corresponds with the demodulation method based on the conventional CMOS demodulation pixel. Since the status of electron storages in the demodulation pixel is changed with an on-off control, the demodulation signal of conventional ToF cameras are the same with the square waveform, as shown in Fig. 4(c) [22,23,27,36,37]. The peak-to-peak value of this signal is also set to exactly 1. According to Fig. 4(c), the ratio of the spectrum intensity at the modulation frequency of 31.25 MHz to that at 0 MHz is about 0.623, which matches the results of previous research works [27]. This tendency is due to the attenuated intensity of the FFT spectrum at 31.25 MHz, which is about 0.3119. Consequently, the amplitude of cross correlation using the square waveform demodulation signal is also decreased compared to the sinusoidal demodulation signal case. According to Eq. (6), the demodulation contrast in this case is about 0.5736, which is quite close to the previous research results [22,25,27].

By analyzing each signal and FFT spectrum, it is shown that digital-parallel demodulation overwhelms other ToF cameras in terms of the demodulation contrast due to the zero-offset and non-attenuated amplitude of cross correlation.

This extremely efficient demodulation characteristic directly improves the measurement precision, even if the integration time is less than 1 microsecond, which is experimentally validated in Section 4.

**3.3 Shot noise-limited distance measurement noise model**

Random noises, such as photon shot noise, dark current noise, and thermal noise, inevitably affect the distance measurement precision of ToF cameras [18,25,27]. These random noises also exist in the proposed AMCW ToF scanning sensor platform as APD is used to detect the laser signal. Based on this fact, the distance measurement noise model of the proposed sensor platform can be expressed following the results of other previous works, as follows [18,27,35]:

$$\Delta L = \left(\frac{c}{4f}\right) \cdot \left(\frac{1}{\sqrt{8}}\right) \cdot \frac{\sqrt{\hat{B} + N_{psuedo}}}{\hat{A}} \tag{7}$$

where $\Delta L$ is the distance measurement noise which is same as the estimated standard deviation of measured distances, $\hat{A}$ is the number of photoelectrons generated by the amplitude of the modulated light source which is proportional to the amplitude of cross correlated samples, $\hat{B}$ is the number of photoelectrons generated by the offset of the modulated light source and ambient light, which is proportional to the offset of cross correlated samples, and $N_{psuedo}$ is the number of electrons which are generated by dark current noise, thermal noise, etc.

For the simplicity of model expression, all digitized signals are assumed to be sinusoidal waveforms. The received laser signal and demodulation signal, which are digitized as shown in Fig. 2, are expressed as follows:

$$r(t) = R \cdot \sin(2\pi ft - \varphi) + R_{DC} \tag{8}$$

$$s(t) = M \cdot \sin(2\pi ft + \varphi_n) + M_{DC}, \quad n = 0,...,3 \tag{9}$$

where $r(t)$ is the received laser signal, $s(t)$ is the demodulation signal, $R$ is the amplitude of the received laser signal, $R_{DC}$ is the offset of the received laser signal, $M$ is the amplitude of the demodulation signal, and $M_{DC}$ is the offset of the demodulation signal.

Using the above digitized signals, the cross correlated sample can be expressed as follows:

$$C(\varphi_n) = \frac{1}{T_{int}} \int_0^{T_{int}} \{(R \cdot \sin(2\pi ft - \varphi) + R_{DC}) \cdot (M \cdot \sin(2\pi ft + \varphi_n) + M_{DC})\} dt \tag{10}$$

where $T_{int}$ is the integration time of the digital-parallel demodulation process.

From mathematical properties of trigonometric functions, the above cross correlation can be expressed as follows:

$$\begin{aligned} C(\varphi_n) = \frac{R \cdot M}{2} \cos(\varphi - \varphi_n) + R_{DC} \cdot M_{DC} + \int_0^{T_{int}} \{R \cdot M_{DC} \sin(2\pi ft - \varphi)\} dt \\ + \int_0^{T_{int}} \{R_{DC} \cdot M \cdot \sin(2\pi ft - \varphi_n)\} dt - \int_0^{T_{int}} \{\frac{R \cdot M}{2} \cdot \cos(4\pi ft - \varphi - \varphi_n)\} dt \end{aligned} \tag{11}$$

If $T_{int}$ is fixed as the integer multiple of the demodulation signal period, the integral terms in Eq. (11) are eliminated. Using this property, cross correlated samples and related parameters are given as follows:

$$C(\varphi_n) = \frac{R \cdot M}{2} \cdot \cos(\varphi - \varphi_n) + R_{DC} \cdot M_{DC} \tag{12}$$

$$A = \frac{\sqrt{\{C(\varphi_0) - C(\varphi_2)\}^2 + \{C(\varphi_3) - C(\varphi_1)\}^2}}{2} = \frac{R \cdot M}{2} \quad (13)$$

$$B = \frac{C(\varphi_0) + C(\varphi_1) + C(\varphi_2) + C(\varphi_3)}{4} = R_{DC} \cdot M_{DC} \quad (14)$$

Meanwhile, the number of electrons in Eq. (7) has a physical relation with the total received energy by APD, energy of one photon, and quantum efficiency, as follows [27]:

$$\hat{A} = P_{amp} \cdot T_{int} \cdot \left(\frac{hc}{\lambda}\right)^{-1} \cdot QE(\lambda) \quad (15)$$

$$\hat{B} = P_{offset} \cdot T_{int} \cdot \left(\frac{hc}{\lambda}\right)^{-1} \cdot QE(\lambda) \quad (16)$$

where $P_{amp}$ is the average received power generated by the amplitude of the modulated laser, $P_{offset}$ is the average received power generated by the offset of the modulated laser and other noise sources, $h$ is the Planck constant, $\lambda$ is the wavelength of the modulated laser whose value is 852 nm, and $QE(\lambda)$ is the quantum efficiency of APD.

The average power of both Eqs. (15) and (16) can be replaced by the parameters related to the assumed signal model and APD parameters as follows [39]:

$$P_{amp} = \frac{R}{R_M(\lambda) \cdot G} \quad (17)$$

$$P_{offset} = \frac{R_{DC}}{R_M(\lambda) \cdot G} \quad (18)$$

where $R_M(\lambda)$ is the responsivity of APD with a value of 23 A/W at the wavelength of 850 nm, $G$ is the transimpedance gain of the amplifier in the APD module with a value of 100 kV/A.

By substituting Eqs. (15), (16), (17), (18) into Eq. (7), Eq. (7) is alternatively expressed as follows:

$$\Delta L = \left(\frac{c}{4f}\right) \cdot \left(\frac{1}{\sqrt{8}}\right) \cdot \frac{\sqrt{R_{DC} \cdot R_M(\lambda) \cdot G \cdot hc}}{R\sqrt{T_{int} \cdot \lambda \cdot QE(\lambda)}} \quad (19)$$

In Eq. (19), the responsivity of APD can be substituted by the quantum efficiency using the following [40]:

$$R_M(\lambda) = \bar{M} \cdot QE(\lambda) \cdot \frac{q\lambda}{hc} \quad (20)$$

where $\bar{M}$ is the multiplication gain of APD with a value of 50 and $q$ is the elementary charge of the electron.

By combining Eq. (19) with Eqs. (13), (14) and (20), the shot noise-limited measurement precision model can be given in terms of the amplitude and offset of the cross correlated samples as follows:

$$\Delta L = \left(\frac{c}{4f}\right) \cdot \frac{M\sqrt{G \cdot q \cdot \bar{M}}}{2\sqrt{T_{int} \cdot M_{DC}}} \cdot \left(\frac{A}{\sqrt{B/2}}\right)^{-1} \quad (21)$$

where $M$ is fixed as 0.4704 in voltage and $M_{DC}$ is fixed as 0.0099 in voltage, for this study, using the demodulation signal shown in Fig. 4(b). Other parameters are explained in the above equations. According to Eq. (21), the distance measurement precision can be estimated if the integration time, the amplitude and offset of the cross correlated samples are known. Uncertainties would exist in Eq. (21) since digitized signals are assumed in ideal sinusoidal waveforms among other unconsidered parameters; however, as experimental conditions become closer to the shot noise-limited case, the estimation of measurement precision based on this model also becomes much more accurate. This tendency is validated in the experimental results shown in the next section.

## 4. EXPERIMENTAL VALIDATION OF AMCW TOF SCANNING SENSOR PLATFORM

### 4.1 Analysis of distance measurement performance

To analyze the distance measurement precision of the proposed AMCW ToF scanning sensor platform, 2,000 data samples were acquired for each reference distance using the reflectance target [41]. Each data sample consists of the measured distance, the amplitude and offset of cross correlated samples. For all experimental conditions in this work, some factors are fixed as follows: the optical illumination power of the laser diode as 30 mW, the frequency of the modulation and demodulation signals as 31.25 MHz, and the environment of measurement condition as bright room (300 lux). Other factors, such as the integration time and measured objects, are changed based on each purpose of the experimental situation. The actual sensor platform and measured reflectance targets are shown in Fig. 5.

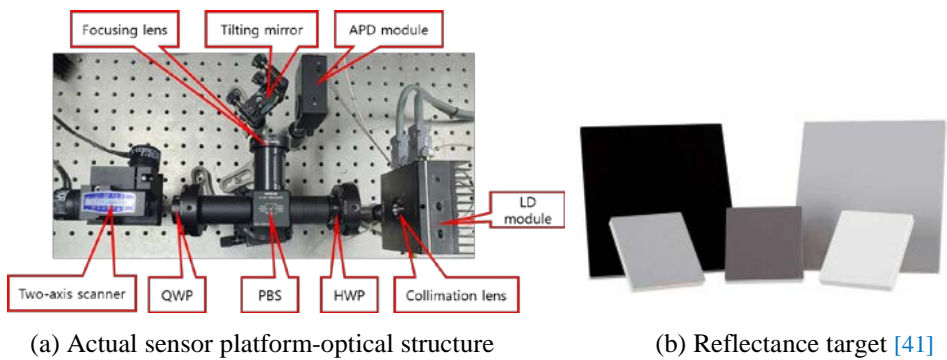

(a) Actual sensor platform-optical structure          (b) Reflectance target [41]

**Fig. 5.** Actual sensor platform and Lambertian reflectance target

Using the sensor platform and reflectance targets shown in Fig. 5, a data set of 2,000 samples measured at specific distance using 95 % reflectance target is shown as an example in Fig. 6.

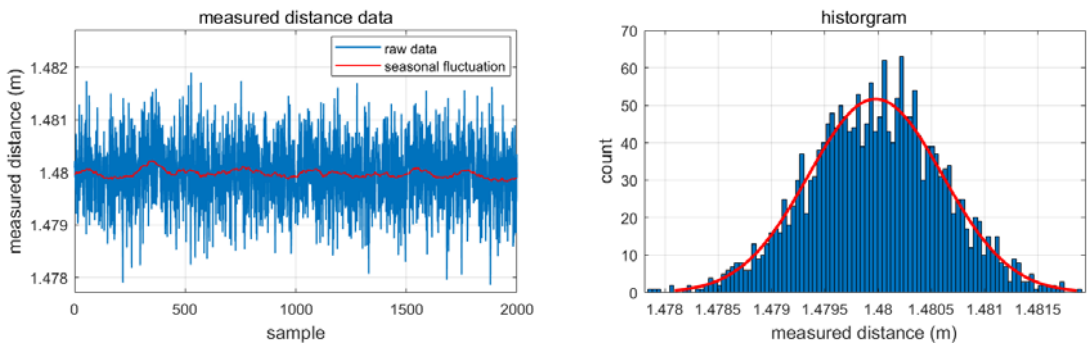

(a) Measured distance

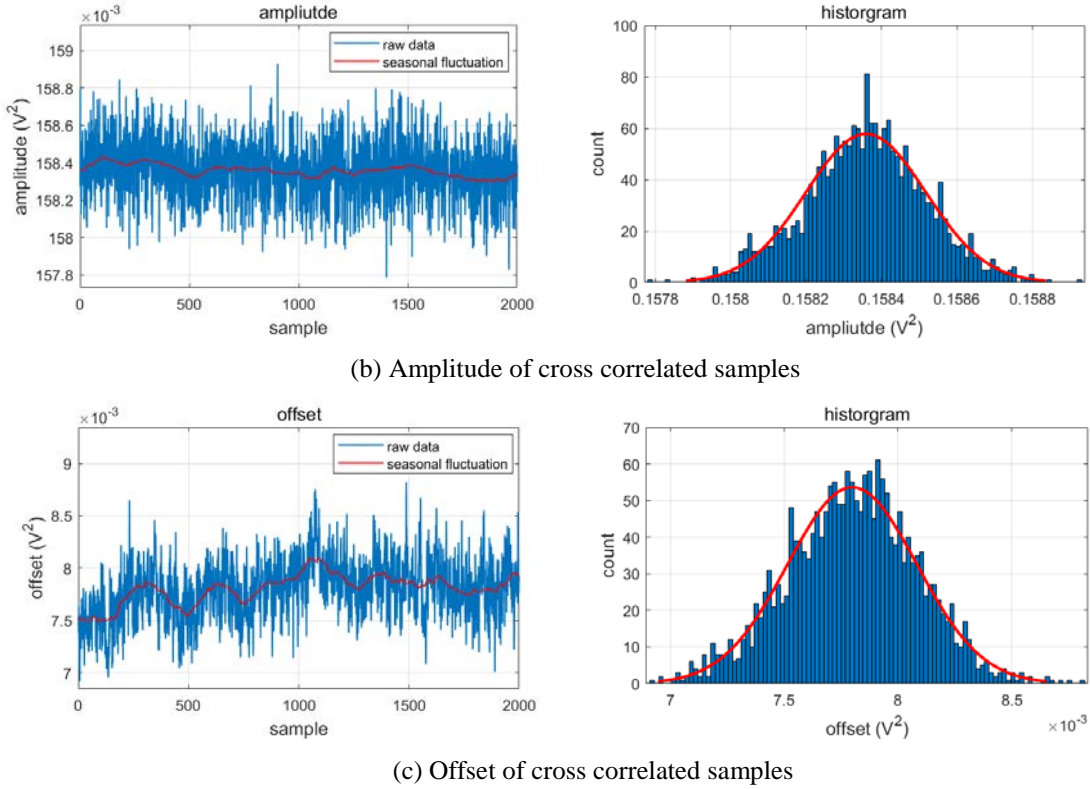

(b) Amplitude of cross correlated samples

(c) Offset of cross correlated samples

**Fig. 6.** Data samples of the measured distance, the amplitude and offset of cross correlated samples at the distance of 1.5 m using 95 % reflectance target.

As shown in Fig. 6, 2,000 data samples are arranged in the order of acquired time, as shown on the left-hand side of the figure, and each corresponding histogram is shown on the right-hand side of the figure. The experimental conditions shown in Fig. 6 are set as follows: the integration time as 16 μsec, a reference distance of 1.5 m, and the reflectivity of the target as 95 %. According to Fig. 6, the distribution of 2,000 data samples for each measured distance, amplitude and offset of cross correlated samples fits well with Gaussian distribution, which is reasonable according to previous research results [35]. In the plots of the sample series of data in Fig. 6, there are temporal deviations for each data point, which are mainly affected by the photon shot noise, dark current noise and laser modulation instability [14,17]. Some seasonal fluctuations, indicated by the red lines in the data sample plots of Fig. 6, exist due to the fluctuation of laser modulation and external disturbances, such as ambient light [14,17]. These seasonal fluctuations are expressed using the moving average filter of which the size is 100 in this work. However, as shown in the measured distance data, the standard deviation of seasonal fluctuation is about 0.072 mm, which is much less than that of the measured distance whose value is approximately 0.628 mm. Meanwhile, the average measured distance is about 1.48 m, which has a difference of 2 cm compared to the reference distance. This measurement error is mainly contributed to the ambient light, stray light, and surface property of the measured object [14,17,20,42]. Such measurement error is generally compensated for by using a look-up table according to other research works [14,17,42]. In this work, the main analysis target is confined to the random noises of the proposed sensor platform, which are directly related to the measurement precision.

Following the experimental process shown in Fig. 6, physically important indices related to the measurement precision, such as the standard deviation of the measured distance, the average amplitude and offset of cross correlated samples are analyzed under various experimental conditions. The experimental conditions are controlled by the reflectivity of the target and the integration time of one distance measurement. The analysis of measured data with a different reflectance target is shown in Fig. 7

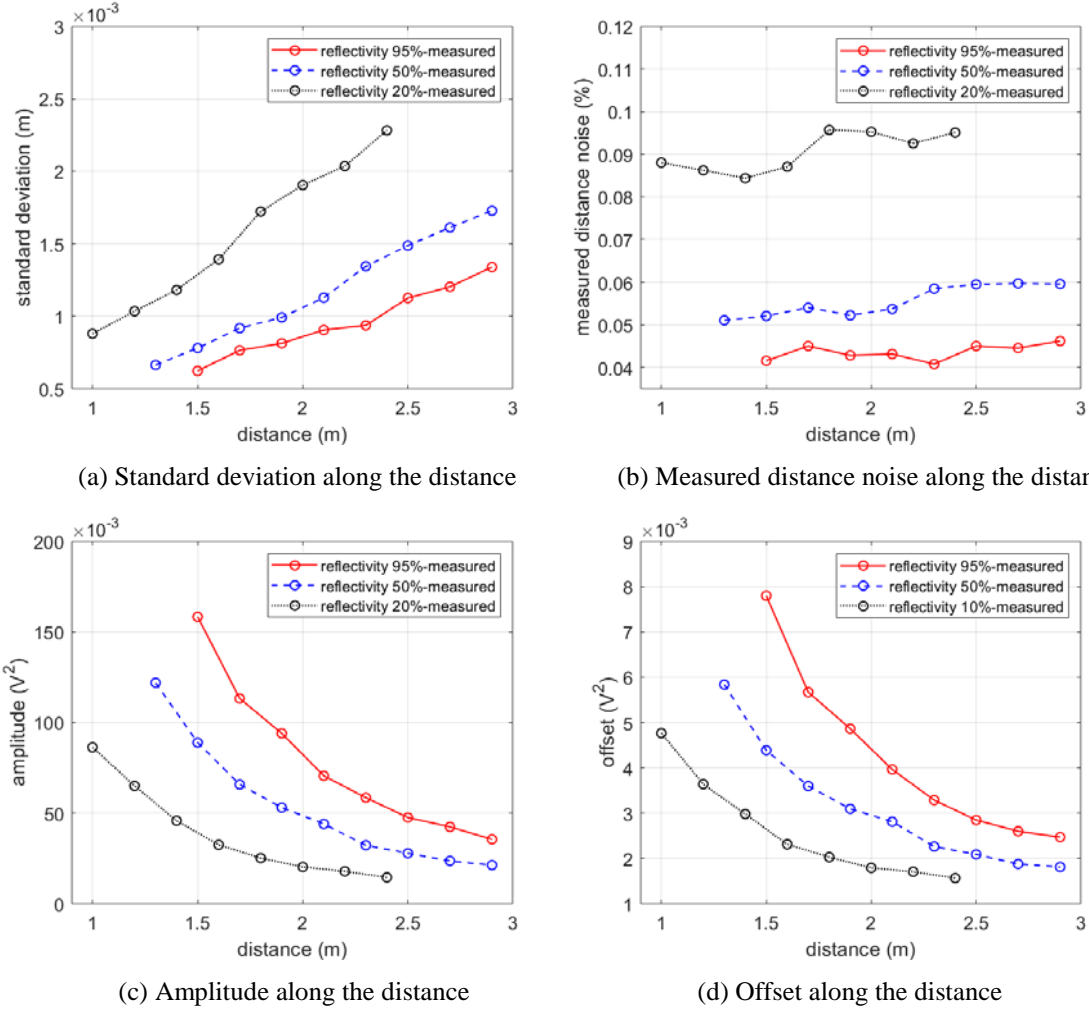

(a) Standard deviation along the distance  (b) Measured distance noise along the distance

(c) Amplitude along the distance  (d) Offset along the distance

**Fig. 7.** Measured data (standard deviation, measured distance noise, amplitude and offset of cross correlated samples) along the reflectance distance using 95 %, 50 %, and 20 % reflectance targets with the integration time of 16 μsec.

As shown in Fig. 7, the experimental conditions are fixed as follows: the integration time per one distance measurement as 16 μsec, measurement range up to 3 m with an interval of 20 cm, target reflectivity of 20 %, 50 % and 95 % with Lambertian surface. For the 20 % reflectance target, the data is acquired up to 2.5 m due to the low received laser power. For 95 % and 50 % reflectance targets, different minimum measurement ranges are set due to the saturation of APD. According to Fig. 7(a), the standard deviation has a tendency to increase monotonically along the reference distance regardless of the reflectivity of the target, which is also shown in other ToF cameras [16,22,27]. The maximum standard deviation of each 95 % and 50 % reflectance target is about 1.3 mm and 1.7 mm, respectively, at the reference distance of 2.9 m. For the target of 20 % reflectance, the maximum standard deviation is about 2.28 mm at the reference distance of 2.4 m. Compared to the standard deviation, the amplitude and offset of cross correlated samples are decreased along the reference distance. The minimum amplitude is 0.035, 0.021 and 0.014 in the dimension of square of voltage for each reflectance target of 95 %, 50 %, and 20 %, respectively. Meanwhile, the offset value is extremely low compared to the amplitude due to the digital-parallel demodulation, as explained in Section 3.2. Except for these indices, measured distance noise is also given, which is defined as follows [37,43]:

$$\delta = \frac{\Delta L_m}{L} \cdot 100 \ (\%) \tag{22}$$

where $\delta$ is the measured distance noise in percentage, $L$ is the reference distance, and $\Delta L_m$ is the standard deviation of the measured distance. According to Fig. 7(b), the maximum value of the measured distance noise is 0.046 %, 0.059 % and 0.095 % for each reflectance target of 95 %, 50 % and 20 %, respectively. Consequently, Fig. 7(b) shows that, as the reflectivity of the target increases, the measurement precision is improved since the amount of received laser is increased.

In addition to changing the reflectivity of the object target, the effects of integration time per one distance measurement were also analyzed, as shown in Fig. 8.

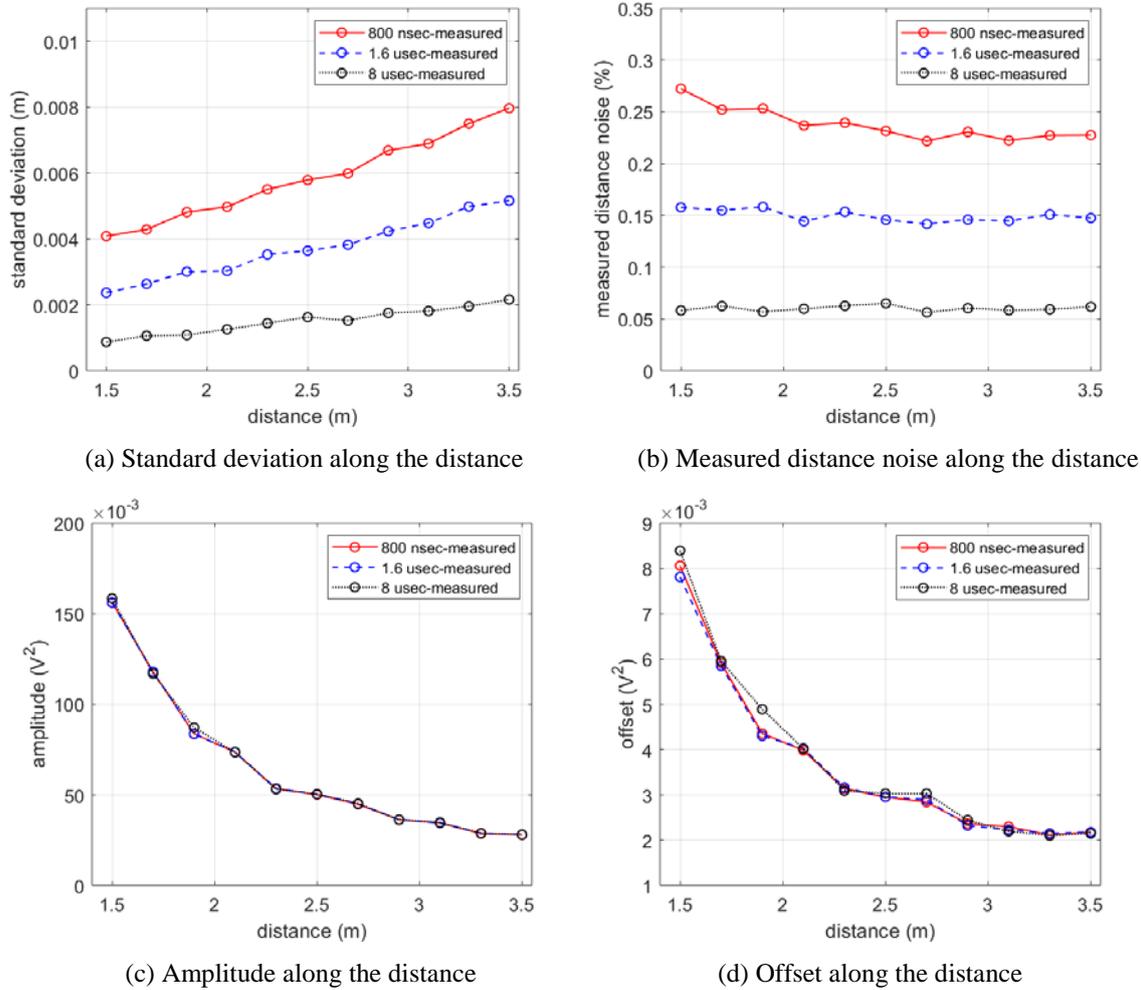

(a) Standard deviation along the distance  (b) Measured distance noise along the distance

(c) Amplitude along the distance  (d) Offset along the distance

**Fig. 8.** Measured data (standard deviation, measured distance noise, amplitude and offset of cross correlated samples) along the reference distance using a 95 % reflectance target with different integration times of 800 nsec, 1.6 μsec, and 8 μsec for each.

As shown in Fig. 8, the experimental conditions are fixed as follows: integration time per one distance measurement as 800 nsec, 1.6 μsec, and 8 μsec, the measurement ranges from 1.5 m to 3.5 m, and the reflectance target fixed as 95 %. Other conditions are the same as previous experiments. According to Fig. 8(a), the maximum value of standard deviation is 7.96 mm, 5.16 mm and 2.16 mm for each integration time of 800 nsec, 1.6 μsec and 8 μsec, respectively, at the reference distance of 3.5 m. The tendency of variation of each physical index along the reference distance is similar with that of Fig. 7. However, the variation of each amplitude and offset of cross correlated samples along the integration time is extremely low compared to the different reflectivity case. Regardless of the value of integration time, the integration term in Eq. (10) is averaged by the length of integration time, which results in nearly the same correlation results. Consequently, there is a relatively small deviation of amplitude and offset at a specific reference distance due to the other error sources, such as the seasonal fluctuation of amplitude and offset, which are shown in Fig. 6. In Fig. 8(b), the measured distance noise defined in Eq. (22) is also given. The maximum measured distance noise is 0.272 %, 0.158 %, and 0.065 % for each integration time of 800 nsec, 1.6 μsec and 8 μsec, respectively. According to the results in Fig. 8, the proposed AMCW ToF sensor platform shows a relatively high precision of distance measurement with extremely low integration time, whose value is less than 1 μsec. Other conventional ToF cameras have a measurement distance noise and integration time that is usually larger than 0.5 % and tens of milliseconds [19,27].

Meanwhile, the comparison of the real standard deviation of measured distance and the measurement precision model defined in Eq. (21) is shown in Fig. 9

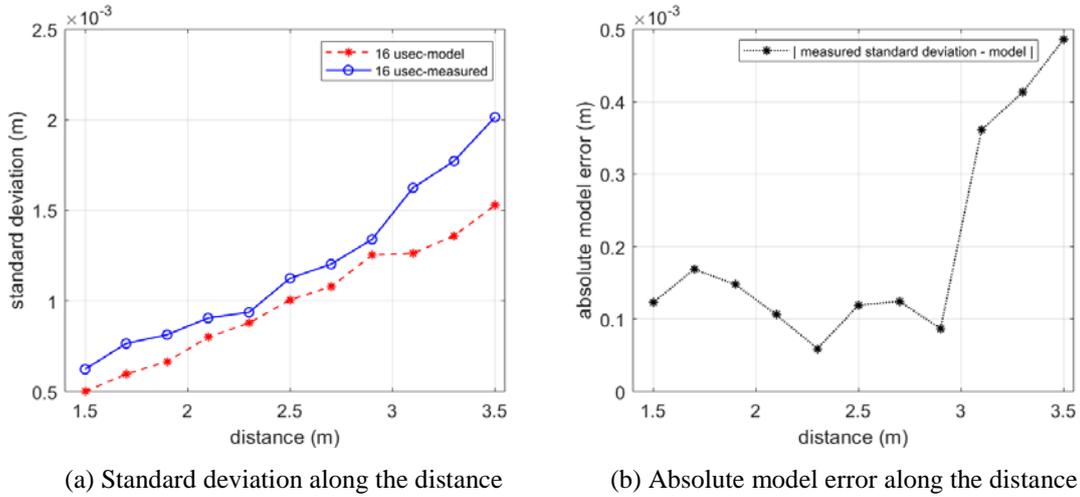

(a) Standard deviation along the distance      (b) Absolute model error along the distance

**Fig. 9.** Comparison of the measured standard deviation and the measurement precision model in Eq. (21)

As shown in Fig. 9, the measured standard deviation and the value of the corresponding model in Eq. (21) are given along the reference distance using the target of 95 % reflectivity. The integration time is fixed as 16 μsec, as shown in Fig. 9, and both the measured standard deviation and corresponding model's value monotonically increase from 1.5 m to 3.5 m. Meanwhile, the absolute model error, which is shown in Fig. 9(b), is defined as the absolute difference between the measured standard deviation and the model. According to Fig. 9(b), at the region of reference distance above 2.7 m, the absolute model error shows a tendency to increase. This tendency can be explained mainly by the discrepancy between the signal model in Eq. (8) and the real digitized signal waveforms. Unlike the assumption of the signal model in Eq. (8), there are other external disturbances, such as stray light from the optical lens and ambient light from external illumination sources, which results in distorted pseudo sinusoidal signals received by APD. As the reference distance increases, the ratio of such external disturbances to the amplitude of the received laser signal increases, which results in an increased discrepancy of the measurement precision model. However, the maximum value of the absolute model error is less than 0.50 mm, which still justifies the measurement precision model in Eq. (21). Meanwhile, the maximum absolute error (MAE) of the precision model shown in Fig. 9 is about 0.1996 mm, which also shows the validity of the measurement precision model in Eq. (21). This validity means that the measurement characteristic of the proposed AMCW ToF scanning sensor in this work is quite close to the shot noise-limited condition.

### 4.2 Comparison of measurement precision with conventional ToF cameras

To compare the measurement precision of the proposed AMCW ToF scanning sensor platform to other existing ToF cameras, a figure of merit (FoM) is defined based on the related works done by Kim et al [36] and Keel et al [37] as follows:

$$\text{FoM} = \frac{\text{illumination power} \times \text{frame acquisition time}}{\text{pixel number}} \times \text{measured distance noise} \quad (23)$$

where frame acquisition time is the consumed time per one image frame acquisition which is the same as the inverse of the frame rate. The measured distance noise is defined in Eq. (22). The product of optical illumination power and frame acquisition time is the same as the total illuminated optical energy per one image frame. By dividing this total illuminated optical energy by the number of pixels in frame, the average illuminated optical energy per one pixel during one image frame acquisition is given. This average illuminated optical energy per one pixel is multiplied by the measured distance noise to define FoM in Eq. (23). This FoM is used to compare the proposed AMCW ToF sensor platform to other widely used existing ToF cameras, as shown in Table 1 [19,20,22,23,28,36,37,44].

**Table 1**
Comparison of the proposed sensor platform to other existing ToF cameras

| ToF sensor model | Image resolution | Frame acquisition time (1/frame rate, sec) | Wavelength (nm) | Illumination power (mW) | Modulation frequency (MHz) | Measurement precision (mm) / distance noise (%) | FoM (nJ/pixel) |
|---|---|---|---|---|---|---|---|
| SR-3000 [20] | 176 × 144 | 0.04 | 850 | 800 (assumption) | 20 | 19 / 2.375 @ 0.8 m | 2998 |
| SR-4000 [19] | 176 × 144 | 0.0185 | 850 | 800 (assumption) | 15, 30 | 8 / 0.5 @ 1.60 m | 291.98 |
| PMD[vision]-19k [19,22] | 160 × 120 | 0.0667 | 870 | 4000 | 20 | 7.5 / 0.5 @ 1.50 m | 6947 |
| PMD-camcube 3.0 [19] | 200 × 200 | 0.0667 | 870 | 800 (assumption) | 21 | 8 / 0.5 @ 1.60 m | 667 |
| PMD-camera module [23] | 172 × 224 | 0.01 | 940, 850 | 1000 | Not specified | 1.5 / 0.15 @ 1 m | 38.933 |
| Kinect V2 [28,44] | 512 × 424 | 0.0333 | 850 | 1000 | multiple | 2.18 / 0.145 @ 1.50 m | 22.24 |
| Kim et al [36] | 320 × 240 | 0.0167 | 855 | 1340 | 10-100 | 0.54 % @ 0.75-4 m | 157.35 |
| Keel et al [37] | 640 × 480 | 0.0167 | 940 | 2000 | 10-150 | 0.30 % @ 1.50 m | 32.617 |
| This work | 320 × 240 | 0.614 | 852 | 30 | 31.25 | 0.84 / 0.056 @ 1.50 m | 13.44 |

The parameters of the proposed sensor platform in this paper, such as image resolution and integration time, are variable. These variable parameters are fixed with reasonable values considering the speed limits of the two-axis scanner used in this paper. The image resolution is fixed as QVGA (320 × 240), and the integration time per pixel is fixed as 8 μsec so that the frame acquisition time (product of the number of pixels and the integration time per pixel) is fixed as 0.614 sec. Meanwhile, for each ToF sensor model, the illumination power is presented based on the other research works [19,20,22,23,26,28,30,31,36,37,44]. Since there is no information on the illumination power for the PMD- camcube 3.0 model, the illumination power for this model is assumed as 800 mW, which is the lowest power compared to other ToF cameras. For the SR series in Table 1, the illumination power is also assumed as 800 mW which is reasonable value according to other previous research results [26,30,31]. Other physical parameters are presented in Table 1. According to Table 1, the FoM of this work is the lowest compared to those of other ToF cameras. This means that, for the same measurement precision, the required optical illumination energy per one image pixel of the sensor platform in this work is lowest compared to that of other ToF cameras. Namely, if the optical illumination energy per one image pixel is the same for all ToF sensor models in Table 1, the proposed sensor platform shows the most precise measurement performance. Such highly efficient measurement performance is mainly caused by the extremely high demodulation contrast and high received optical SNR per one image pixel of the proposed sensor platform, which have been previously explained in Section 3.1 and 3.2.

### 4.3 Analysis of 3D depth images

3D depth maps and corresponding amplitude maps are captured under various experimental conditions to demonstrate the measurement performance of the proposed AMCW ToF scanning sensor platform. Before analyzing a complex scene with multi objects, a single object is captured and analyzed. A raw 3D depth point cloud of a single object and the related measurement error are shown in Fig. 10.

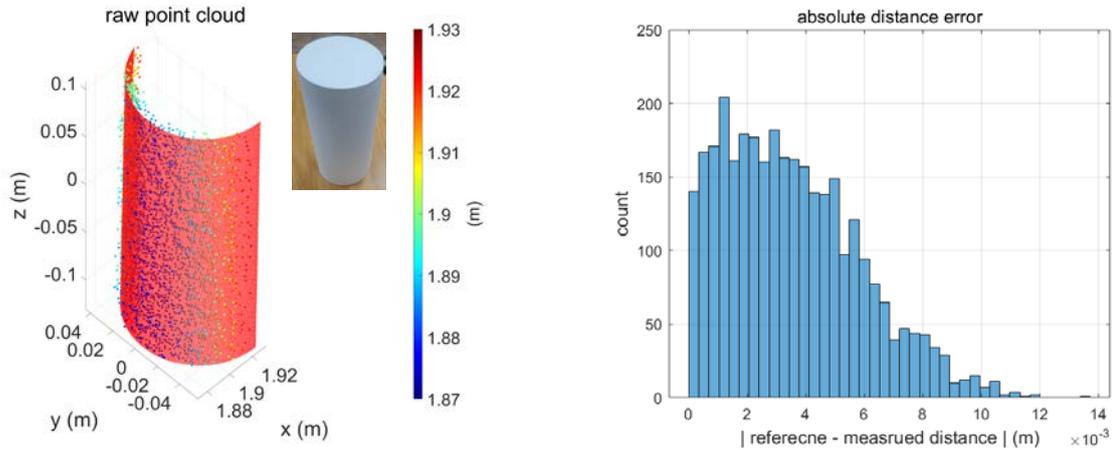

(a) Measured point cloud and the reference model (red cylinder)  (b) Histogram of absolute distance error

**Fig. 10.** Measured point cloud (colored points) of reference model (red cylinder) and the histogram of absolute distance errors.

The red cylinder in Fig. 10(a) is the point cloud of the reference model which is a cylinder with a radius of 5 cm located at the distance of about 1.88 m. The reference model of the cylinder generated by simulation is aligned with the measured point cloud using the Iterative Closest Point (ICP) algorithm [16]. There are 40,000 measured points in the point cloud, as shown in Fig. 10(a), as the image resolution is set as 200 by 200 in this measurement condition. The integration time per one image pixel in this experiment is set as 800 nsec. For each measured point, the absolute distance between the measured point and corresponding reference model point is calculated and shown as the histogram in Fig. 10(b) [16]. This histogram shows that the maximum value of the absolute distance error is less than 14 mm, which is much smaller than that of Kinect V2 which is larger than 40 mm according to the work of A.Corti et al [16]. Consequently, the proposed AMCW ToF scanning sensor platform shows an improved measurement accuracy compared to the Kinect V2, although the distance of the object in this experiment is longer than that shown in the work of A.Corti et al [16].

Meanwhile, to analyze the complex scene of multi-objects using the proposed AMCW ToF scanning sensor platform, multiple 3D depth maps and corresponding amplitude maps are captured. One of the multi-object scenes is shown in Fig. 11.

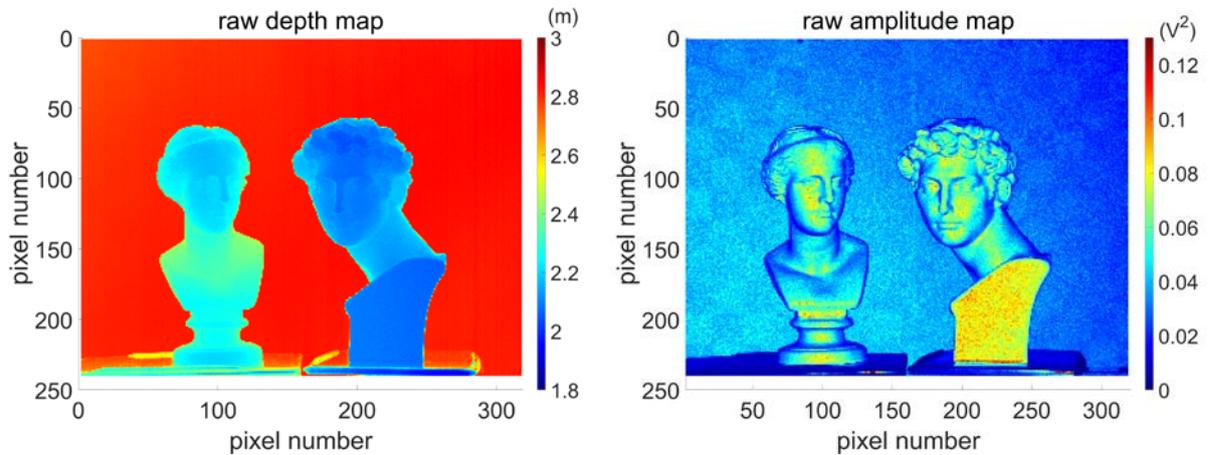

(a) Raw depth map- frontal view  (b) Raw amplitude map

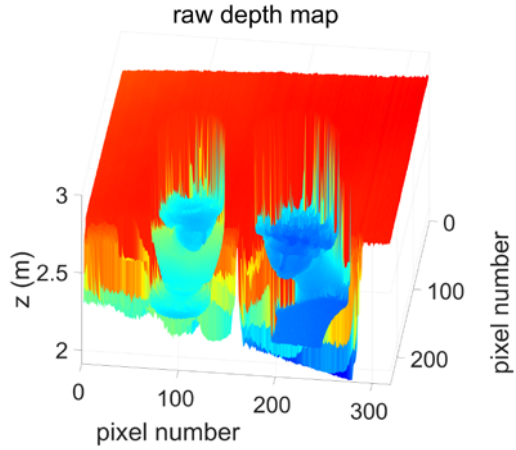
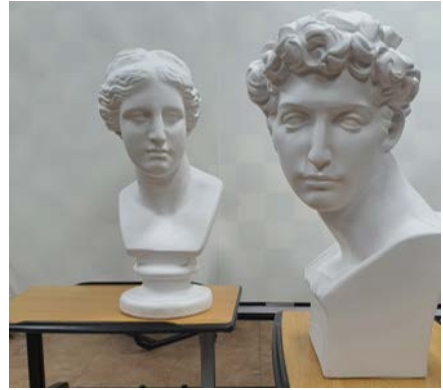

(c) Raw depth map- side view　　　　　　　　(d) Actual object-photograph

**Fig. 11.** 3D depth map, amplitude map, and actual object scene, in QVGA resolution with the integration time of 800 nsec.

All depth and amplitude maps in Fig. 11(a)-(c) are raw images which are not post-processed using spatial filters such as an edge denoising filter [45]. These raw depth and amplitude maps are captured at QVGA resolution with the integration time of 800 nsec per pixel. As shown in Fig. 11(a) and (c), the overall contour and features of busts are well presented in the raw depth map. The plane behind the busts is also well represented as a flat surface, as shown in Fig. 11(a) and (c). In the amplitude map, as the measured object becomes closer to the sensor platform, the amplitude increases as the power of received light is inversely proportional to the square of the object distance [27]. Meanwhile, the checkboard pattern is presented in the amplitude map in Fig. 11(b), which is due to the different reflectivity of surface texture, as shown in the background of the actual object scene in Fig. 11(d). This checkboard pattern is not found in the depth map in Fig. 11(a), which means that the proposed sensor platform has robustness in distance measurement against relatively small amplitude variation due to the reflectivity of the object.

To analyze the 3D depth map and corresponding amplitude map more specifically, two intrinsic parameters: the integration time and the image resolution, are controlled. Specifically, 3D depth and amplitude maps are captured with the same image resolution and measurement object scene for each integration time per pixel: 800 nsec and 16 μsec. Related results are shown in Fig. 12 and 13.

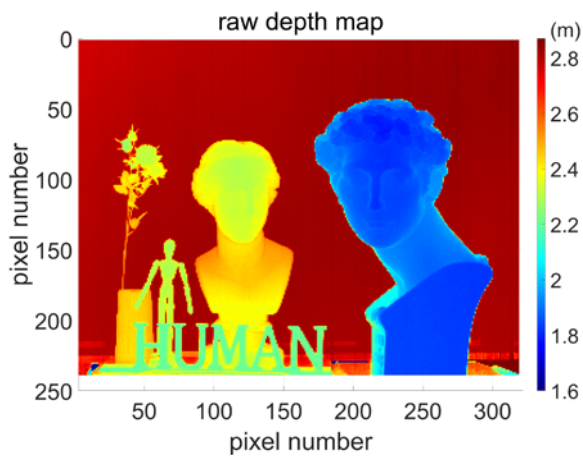
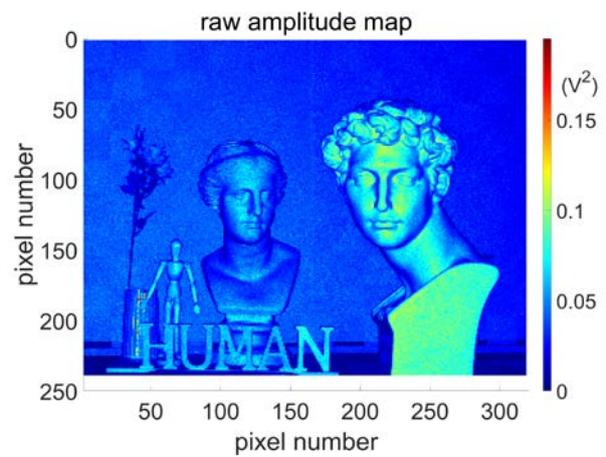

(a) Raw depth map- frontal view　　　　　　　　(b) Raw amplitude map

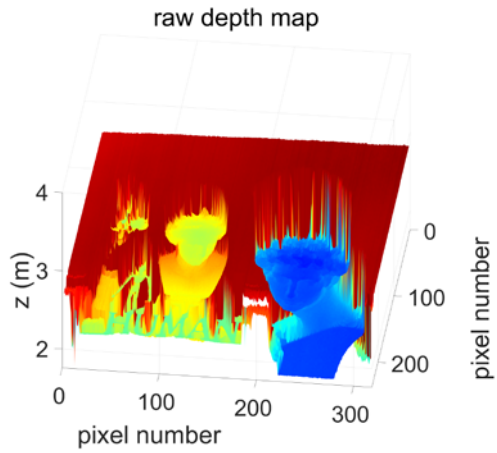
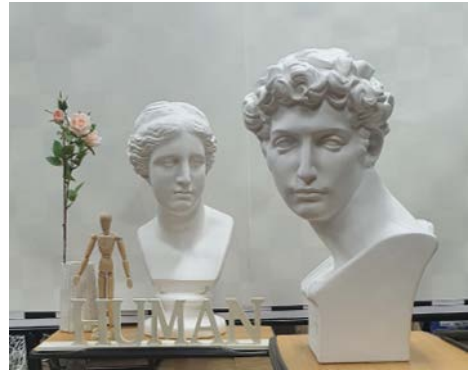

(c) Raw depth map- side view  (d) Actual object-photograph

**Fig. 12.** 3D depth map, amplitude map, and actual object scene, in QVGA resolution with the integration time of 16 μsec.

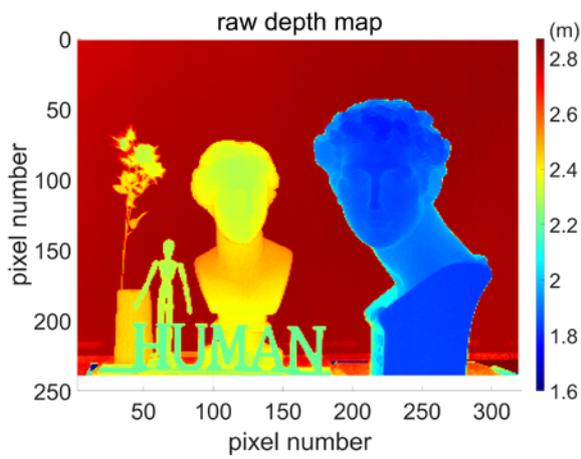
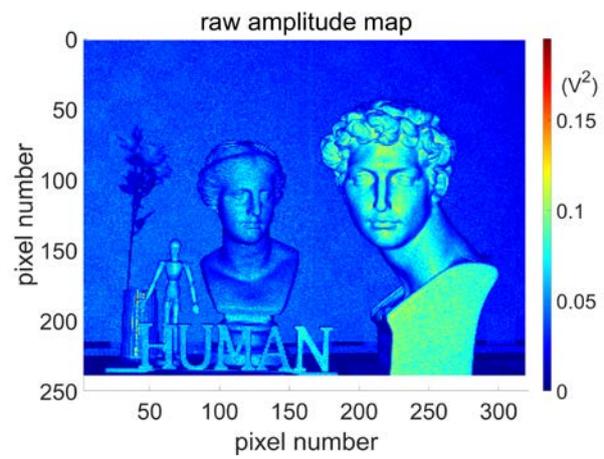

(a) Raw depth map- frontal view  (b) Raw amplitude map

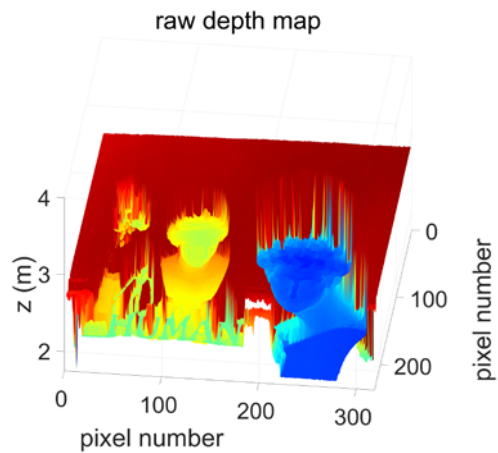

(c) Raw depth map- side view

**Fig. 13.** 3D depth map, amplitude map in QVGA resolution with the integration time of 800 nsec.

Fig. 12 shows the frontal and side view of the 3D depth map, the amplitude map, and the actual object scene. All 3D images are raw data with no processed spatial filter. In the 3D depth map, overall contour and features of busts and other objects, such as background plane, characters, wooden statue, and flowers, are well expressed. One distinguished point compared to the depth map of Fig. 11 is the checkboard pattern at the bottom horizontal line of the background plane in the 3D depth map. This pattern occurs due to the abrupt reflectivity variation of the surface. Unlike other parts of the background plane, the bottom side is not covered by white paper, as shown in the actual object scene photograph in Fig. 12(d), which results in an abrupt reflectivity variation along the horizontal line at the bottom of the background plane. This means that a distance measurement error inevitably occurs if the reflectivity variation exceeds a specific threshold, which is also shown in other widely used ToF cameras [14,17,42]. There is also a phase ambiguity issue in the bottom left side of the frontal view (dark blue) of the 3D depth map in Fig. 12(a). This phase ambiguity also occurs mostly due to the limitation of the single frequency modulation method. To solve this phase ambiguity problem, many related research works have already been presented using multiple frequency modulation, which is beyond the scope of this paper [46]. Except for the amplitude related error and phase ambiguity error, other critical distance measurement errors are not present in Fig. 12(a) and (c). Meanwhile, in the amplitude map, although the leaves and flowers are located close to each other, the amplitude values for each part are quite different. However, the effect of such amplitude difference on the distance measurement is not considerable, as shown in the 3D depth map of Fig. 12(a) and (c), which indicates the robustness of distance measurement of the proposed sensor platform.

The actual object scene shown in Fig. 12(d) is also used to capture 3D images using different integration times, as shown in Fig. 13. The main difference of the 3D depth maps of Fig. 13(a) compared to that of Fig. 12(a) is that there are some blurs at the region of flowers. This blur-like imaging is due to the relatively poor measurement precision as the integration time of Fig. 13 is 800 nsec which is extremely low compared to that of Fig. 12. As integration time decreases, the measurement precision also decreases, as shown in Fig. 8. Although there are some blurs in the flowers, there are no specific problems related to the measurement precision in other regions of the depth map, such as busts, the wooden statue and characters. Other characteristics of the 3D images in Fig. 13 can be explained in the same way as Fig. 12. By comparing each 3D image using different integration time, it is clear that as integration time increases, the 3D images become more precise; however, the processing time per one image frame also becomes longer.

The image resolution is also changed to capture the same object scene, as shown in Fig. 14.

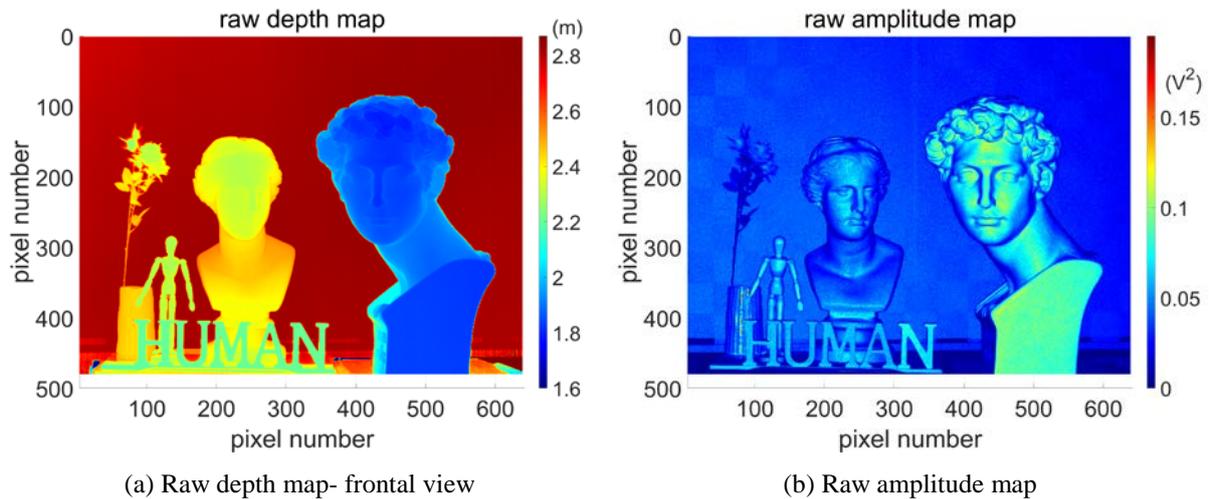

(a) Raw depth map- frontal view  (b) Raw amplitude map

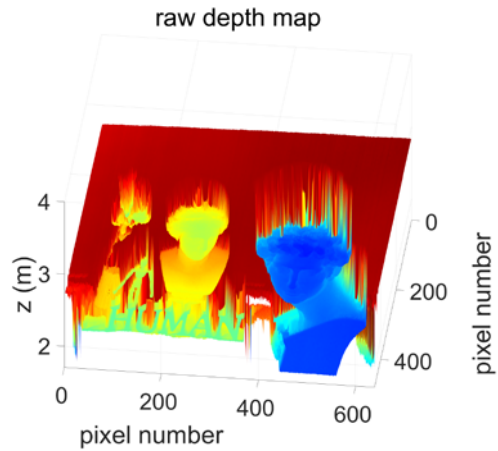

(c) Raw depth map- side view

**Fig. 14.** 3D depth map, amplitude map in VGA resolution with the integration time of 800 nsec.

All measurement conditions in Fig. 14 are the same as Fig. 13 except for the image resolution scale: VGA (640 × 480). As shown in Fig. 14, image quality in the aspect of spatial resolution has been improved and the overall image maps look much clearer compared to the QVGA maps. Volumetric features, such as the nose and eyes of busts and flowers, look much clearer than that of Fig. 13. However, there are still blur-like noises in the region of the flower, which can be improved by increasing the integration time. Meanwhile, as the image resolution increases, the amplitude map also looks sharper compared to that of Fig. 13, although no spatial filter is applied. In contrast, as image resolution increases, the frame rate inevitably decreases. To lower the frame acquisition time per one image frame, the integration time should be decreased, which has a trade-off relationship with the measurement precision. However, in these experiments, the optical power of laser source is only 30 mW, which means that the frame rate can be improved as well maintaining the precision by increasing the optical power of the laser source.

All these image resolutions and integration times per pixel can be flexibly chosen in the proposed sensor platform. This means that an extremely high super image resolution of a 3D depth map is also achievable. Using the proposed sensor platform, the FHD (1920 × 1080) image resolution of the 3D depth map is validated, as shown in Fig. 15.

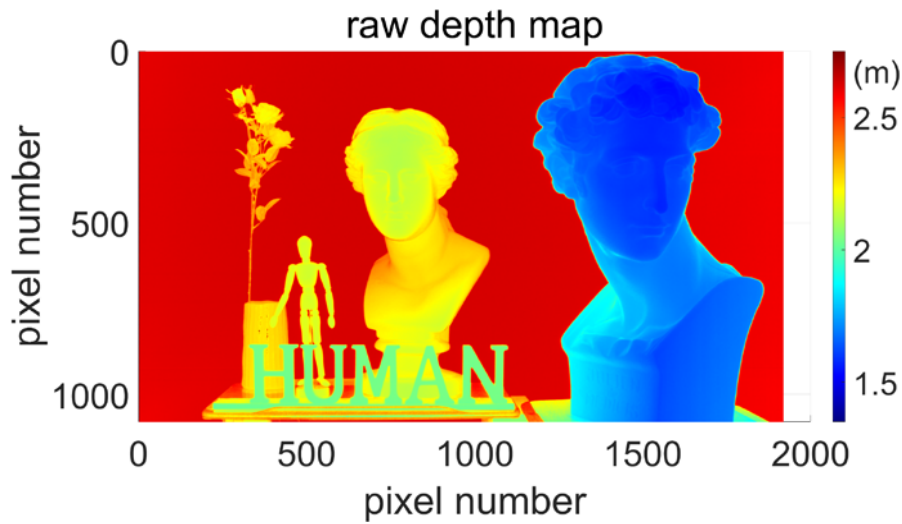

(a) Raw depth map- frontal view

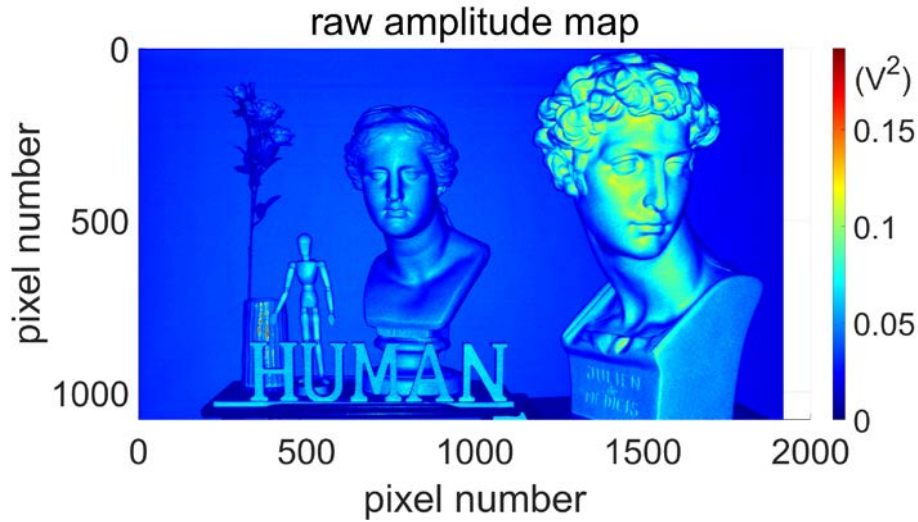

(b) Raw amplitude map

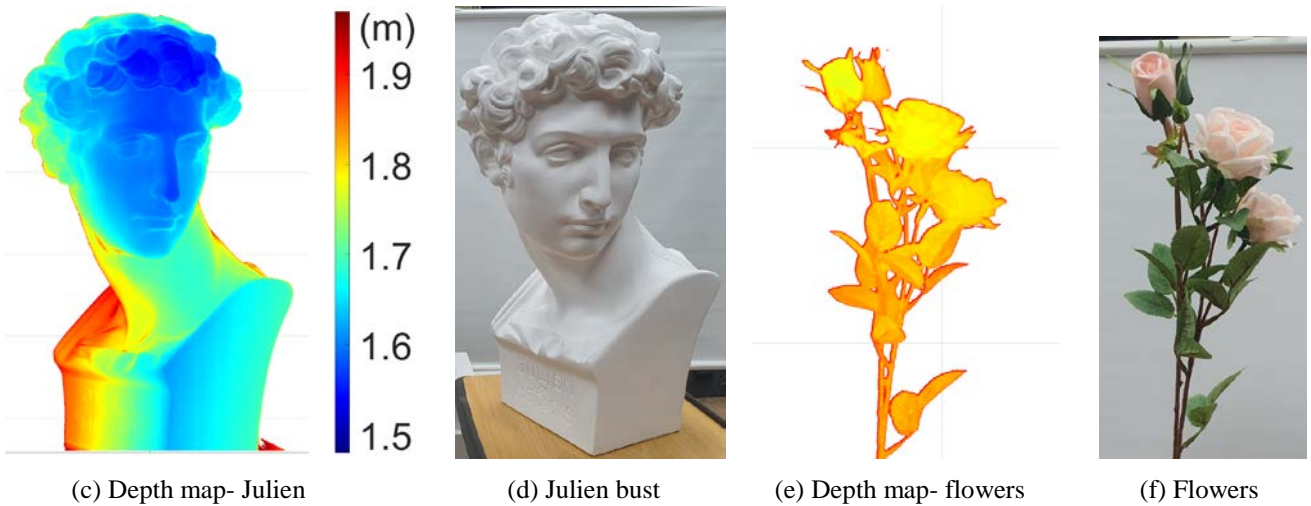

(c) Depth map- Julien    (d) Julien bust    (e) Depth map- flowers    (f) Flowers

**Fig. 15.** 3D depth map, amplitude map in FHD resolution with the integration time of 800 nsec.

In Fig. 15, the integration time per pixel is 800 nsec and the image resolution is in FHD (1920 × 1080) scale for each. The details of the busts are extremely well presented in the 3D depth maps, as shown in Fig. 15(a) and (c). Even the phrase 'JULIEN de MEDICIS' is clearly expressed in the 3D depth map of the bust, as shown in Fig. 15(a) and (c). Additionally, these very high resolution depth images can also express the stem of the leaves and detailed contours of the flowers, as shown in Fig. 15(e). These extremely precise and high resolution depth measurement results are achievable using the AMCW ToF sensor platform proposed in this paper, which is not feasible using other existing ToF cameras whose a maximum image resolution is in VGA scale [37]. These extremely precise, super resolution depth images in Fig. 15 indicate the strong advantages of the proposed sensor platform over other existing ToF cameras.

In summary, by changing the integration time and image resolution, a 3D depth map and corresponding amplitude map are analyzed. As the integration time increases with the same image resolution, 3D images become much more precise; however, the processing time per one image frame increases. Meanwhile, as the image resolution increases, the volumetric features of objects become much clearer as shown in Fig. 16, which also induces an increased frame acquisition time.

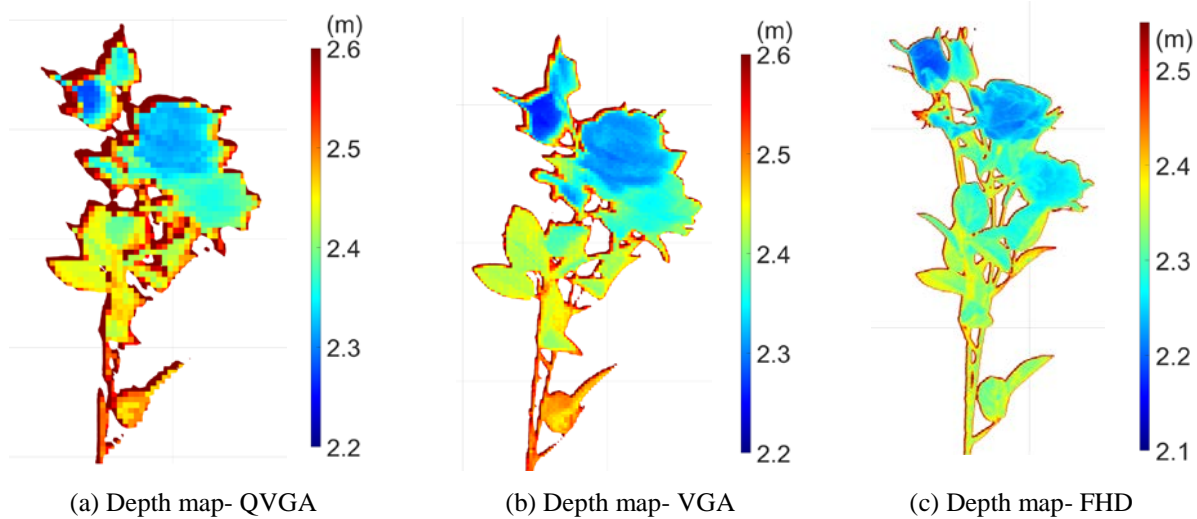

(a) Depth map- QVGA      (b) Depth map- VGA      (c) Depth map- FHD

**Fig. 16.** 3D depth maps of flowers in QVGA, VGA, and FHD resolution with the integration time of 800 nsec.

## CONCLUSION

In this paper, a novel AMCW ToF scanning sensor platform based on digital-parallel demodulation is demonstrated. According to the experimental results, the proposed sensor platform shows the most efficient distance measurement performance compared to other ToF cameras in the aspect of FoM which is defined by the optical illumination energy and measurement precision, as shown in Table 1. Additionally, the extremely precise super resolution 3D depth map is validated, which indicates the superiority of the proposed sensor platform to other ToF cameras in the aspect depth measurement precision. This improved measurement precision is based on the properties of the digital-parallel demodulation method and 2D scanning structure with a single photodetector, which are its primary distinguishing characteristics compared to that of other widely used ToF cameras. To commercialize this proposed sensor platform for practical application, much smaller and faster scanners, such as MEMS scanners, should be utilized in future studies [6,9]. Meanwhile, some issues remain regarding distance measurement accuracy. As shown in Figs. 12, 13 and 14, the abrupt variation of amplitude induces a distance measurement error, which is one of the main research issues related to ToF cameras [14,17,42]. Additionally, phase ambiguity, which is inevitable for single frequency modulation, should also be solved [46]. With the realization of further improvements related to the size and measurement accuracy of the sensor platform, the AMCW ToF scanning sensor based on digital-parallel demodulation proposed in this paper can be used in various industrial and academic applications.

## ACKNOWLEDGEMENTS

This study is a part of the research project, "Development of core machinery technologies for autonomous operation and manufacturing", which has been supported by a grant from National Research Council of Science & Technology under the R&D Program of the Ministry of Science, ICT and Future Planning, and a part of the project titled "Human Resources Program in Energy Technology of the Korea Institute of Energy Technology Evaluation and Planning (KETEP)", funded by the Ministry of Trade, Industry & Energy, Republic of Korea (Grant No. 20204030200050).

## REFERENCES

[1]    H. Gao, B. Cheng, J. Wang, K. Li, J. Zhao, D. Li, Object Classification Using CNN-Based Fusion of Vision and LIDAR in Autonomous Vehicle Environment, IEEE Trans. Ind. Informatics. 14 (2018) 4224–4230. https://doi.org/10.1109/TII.2018.2822828.

[2]  F. Endres, J. Hess, N. Engelhard, J. Sturm, D. Cremers, W. Burgard, An evaluation of the RGB-D SLAM system, Proc. - IEEE Int. Conf. Robot. Autom. 1 (2012) 1691–1696. https://doi.org/10.1109/ICRA.2012.6225199.

[3]  D. Buchholz, M. Futterlieb, S. Winkelbach, F.M. Wahl, Efficient bin-picking and grasp planning based on depth data, Proc. - IEEE Int. Conf. Robot. Autom. (2013) 3245–3250. https://doi.org/10.1109/ICRA.2013.6631029.

[4]  T. Hernández-Díaz, A. Vázquez-Cervantes, J.J. Gonzalez-Barboza, L. Barriga-Rodríguez, A.M. Herrera-Navarro, L.A. Baldenegro-Pérez, H. Jiménez-Hernández, Detecting background and foreground with a laser array system, Meas. J. Int. Meas. Confed. 63 (2015) 195–206. https://doi.org/10.1016/j.measurement.2014.12.004.

[5]  T. Sielhorst, C. Bichlmeier, S.M. Heining, N. Navab, Depth perception - A major issue in medical AR: Evaluation study by twenty surgeons, Lect. Notes Comput. Sci. (Including Subser. Lect. Notes Artif. Intell. Lect. Notes Bioinformatics). 4190 LNCS (2006) 364–372. https://doi.org/10.1007/11866565_45.

[6]  H. Xie, S. Strassle, S. Koppal, A. Stainsby, Y. Bai, D. Wang, A compact 3D lidar based on an electrothermal two-axis MEMS scanner for small UAV, in: Laser Radar Technol. Appl. XXIII, 2018: p. 14. https://doi.org/10.1117/12.2304529.

[7]  T. Raj, F.H. Hashim, A.B. Huddin, M.F. Ibrahim, A. Hussain, A survey on LiDAR scanning mechanisms, Electron. 9 (2020). https://doi.org/10.3390/electronics9050741.

[8]  A. McCarthy, R.J. Collins, N.J. Krichel, V. Fernández, A.M. Wallace, G.S. Buller, Long-range time-of-flight scanning sensor based on high-speed time-correlated single-photon counting, Appl. Opt. 48 (2009) 6241–6251. https://doi.org/10.1364/AO.48.006241.

[9]  P.K. Choudhury, C.H. Lee, Simultaneous Enhancement of Scanning Area and Imaging Speed for a MEMS Mirror Based High Resolution LiDAR, IEEE Access. 8 (2020) 52113–52120. https://doi.org/10.1109/ACCESS.2020.2979326.

[10]  T. Hegna, H. Pettersson, K.M. Laundal, K. Grujic, 3D laser scanner system based on a galvanometer scan head for high temperature applications, Opt. Meas. Syst. Ind. Insp. VII. 8082 (2011) 80823Z. https://doi.org/10.1117/12.888985.

[11]  L. Ye, G. Gu, W. He, H. Dai, Q. Chen, A Real-Time Restraint Method for Range Walk Error in 3-D Imaging Lidar Via Dual Detection, IEEE Photonics J. 10 (2018) 1–9. https://doi.org/10.1109/JPHOT.2018.2816652.

[12]  R. Horaud, M. Hansard, G. Evangelidis, C. Ménier, An overview of depth cameras and range scanners based on time-of-flight technologies, Mach. Vis. Appl. 27 (2016) 1005–1020. https://doi.org/10.1007/s00138-016-0784-4.

[13]  Z. Wang, Review of real-time three-dimensional shape measurement techniques, Meas. J. Int. Meas. Confed. 156 (2020) 107624. https://doi.org/10.1016/j.measurement.2020.107624.

[14]  S. Foix, G. Alenyà, C. Torras, Lock-in time-of-flight (ToF) cameras: A survey, IEEE Sens. J. 11 (2011) 1917–1926. https://doi.org/10.1109/JSEN.2010.2101060.

[15]  T. Oggier, M. Lehmann, R. Kaufmann, M. Schweizer, M. Richter, P. Metzler, G. Lang, F. Lustenberger, N. Blanc, An all-solid-state optical range camera for 3D real-time imaging with sub-centimeter depth resolution (SwissRanger), Opt. Des. Eng. 5249 (2004) 534. https://doi.org/10.1117/12.513307.

[16]  A. Corti, S. Giancola, G. Mainetti, R. Sala, A metrological characterization of the Kinect V2 time-of-flight camera, Rob. Auton. Syst. 75 (2016) 584–594. https://doi.org/10.1016/j.robot.2015.09.024.

[17]  Y. He, S. Chen, Recent Advances in 3D Data Acquisition and Processing by Time-of-Flight Camera, IEEE Access. 7 (2019) 12495–12510. https://doi.org/10.1109/ACCESS.2019.2891693.

[18]  R. Lange, P. Seitz, Solid-state time-of-flight range camera, IEEE J. Quantum Electron. 37 (2001) 390–397. https://doi.org/10.1109/3.910448.

[19]  D. Piatti, F. Rinaudo, SR-4000 and CamCube3.0 Time of Flight (ToF) Cameras: Tests and Comparison, Remote Sens. 4 (2012) 1069–1089. https://doi.org/10.3390/rs4041069.

[20]  P. Fursattel, S. Placht, M. Balda, C. Schaller, H. Hofmann, A. Maier, C. Riess, A Comparative Error Analysis of Current Time-of-Flight Sensors, IEEE Trans. Comput. Imaging. 2 (2015) 27–41. https://doi.org/10.1109/tci.2015.2510506.

[21]  Y.-H. Park, Y.-C. Cho, J.-W. You, C.-Y. Park, H.-S. Yoon, S.-H. Lee, J.-O. Kwon, S.-W. Lee, B.H. Na, G.W. Ju, H.J. Choi, Y.T. Lee, Three-dimensional imaging using fast micromachined electro-absorptive shutter, J. Micro/Nanolithography, MEMS, MOEMS. 12 (2013) 023011. https://doi.org/10.1117/1.jmm.12.2.023011.

[22]  M. Tobias, K. Holger, F. Jochen, A. Martin, L. Robert, Robust 3D Measurement with PMD Sensors, in: Range Imaging Day, Zürich, 2005: p. 8. http://citeseerx.ist.psu.edu/viewdoc/download?doi=10.1.1.132.5821&rep=rep1&type=pdf.

[23]  R. Lange, S. Böhmer, B. Buxbaum, CMOS-based optical time-of-flight 3D imaging and ranging, 2019. https://doi.org/10.1016/B978-0-08-102434-8.00011-8.

[24] M. Bueno, L. Díaz-Vilariño, J. Martínez-Sánchez, H. González-Jorge, H. Lorenzo, P. Arias, Metrological evaluation of KinectFusion and its comparison with Microsoft Kinect sensor, Meas. J. Int. Meas. Confed. 73 (2015) 137–145. https://doi.org/10.1016/j.measurement.2015.05.018.

[25] R. Lange, P. Seitz, A. Biber, S.C. Lauxtermann, Demodulation pixels in CCD and CMOS technologies for time-of-flight ranging, in: Sensors Camera Syst. Sci. Ind. Digit. Photogr. Appl., 2000: p. 177. https://doi.org/10.1117/12.385434.

[26] T. Kahlmann, H. Ingensand, Calibration of the fast range imaging camera SwissRanger for use in the surveillance of the environment, Electro-Optical Remote Sens. II. 6396 (2006) 639605. https://doi.org/10.1117/12.684458.

[27] R. Lange, Time-of-flight distance measurement with with custom custom solid-state image sensors in CMOS / CCD-technology, Sensors Peterbrgh. NH. 222 (2000) 223. http://dokumentix.ub.uni-siegen.de/opus/volltexte/2006/178/.

[28] Z. Xu, T. Perry, G. Hills, Method and system for multi-phase dynamic calibration of three-dimensional (3D) sensors in a time-of-flight system, U.S. Patent No. 8,587,771. (2013).

[29] M. Plaue, Technical Report : Analysis of the PMD Imaging System, Univ. Heidelb. (2006). http://www.researchgate.net/publication/228916096_Technical_Report_Analysis_of_the_PMD_Imaging_System/file/79e4151450a8eec67c.pdf.

[30] D. Anderson, H. Herman, Experimental characterization of commercial flash ladar devices, Int. Conf. Sens. (2005) 3–8. http://citeseerx.ist.psu.edu/viewdoc/download?doi=10.1.1.66.2608&rep=rep1&type=pdf.

[31] H. Rapp, Experimental and theoretical investigation of correlating TOF-camera systems. (2007). https://doi.org/10.11588/heidok.00007666.

[32] M. Imaki, S. Kameyama, E. Ishimura, M. Nakaji, H. Yoshinaga, Y. Hirano, Line scanning time-of-flight laser sensor for intelligent transport systems, combining wide field-of-view optics of 30 deg, high scanning speed of 0.9 ms / line , and simple sensor configuration, Opt. Eng. 56 (2016) 031205. https://doi.org/10.1117/1.oe.56.3.031205.

[33] C. Zhang, N. Hayashi, S.Y. Set, S. Yamashita, Irradiation angle dependence and polarization dependence in 3D geometry measurement using AMCW LiDAR, Laser 3D Manufacturing VI. 10909 (2019) 109090W. https://doi.org/10.1117/12.2507728.

[34] S.H. Lee, W.-H. Kwon, Y.-H. Park, Amplitude-modulated continuous wave scanning LIDAR based on parallel phase-demodulation, MOEMS and Miniaturized Systems XX. 11697 (2021) 116970H. https://doi.org/10.1117/12.2577122.

[35] F. Mufti, R. Mahony, Statistical analysis of signal measurement in time-of-flight cameras, ISPRS J. Photogramm. Remote Sens. 66 (2011) 720–731. https://doi.org/10.1016/j.isprsjprs.2011.06.004.

[36] D. Kim, S. Lee, D. Park, C. Piao, J. Park, Y. Ahn, K. Cho, J. Shin, S.M. Song, S.J. Kim, J.H. Chun, J. Choi, Indirect Time-of-Flight CMOS Image Sensor with On-Chip Background Light Cancelling and Pseudo-Four-Tap/Two-Tap Hybrid Imaging for Motion Artifact Suppression, IEEE J. Solid-State Circuits. 55 (2020) 2849–2865. https://doi.org/10.1109/JSSC.2020.3021246.

[37] M.S. Keel, Y.G. Jin, Y. Kim, D. Kim, Y. Kim, M. Bae, B. Chung, S. Son, H. Kim, T. An, S.H. Choi, T. Jung, Y. Kwon, S. Seo, S.Y. Kim, K. Bae, S.C. Shin, M. Ki, S. Yoo, C.R. Moon, H. Ryu, J. Kim, A VGA Indirect Time-of-Flight CMOS Image Sensor with 4-Tap 7-μm Global-Shutter Pixel and Fixed-Pattern Phase Noise Self-Compensation, IEEE J. Solid-State Circuits. 55 (2020) 889–897. https://doi.org/10.1109/JSSC.2019.2959502.

[38] A.I. Zayed, A convolution and product theorem for the fractional Fourier transform, IEEE Signal Process. Lett. 5 (1998) 101–103. https://doi.org/10.1109/97.664179.

[39] https://www.thorlabs.com/drawings/972b082c1fefbf17-A1DC822F-B68D-CC52- AAC9AC9F1AC9046B/ APD130A_M-Manual.pdf (accessed 19th August 2021).

[40] Safa, O., et al. "OPTOELECTRONICS & PHOTONICS: PRINCIPLES & PRACTICES (2ND." (2012).

[41] https://sphereoptics.de/en/product/zenith-lite-ultralight-targets (accessed 19th August 2021).

[42] M. Lindner, A. Kolb, Calibration of the intensity-related distance error of the PMD TOF-camera, Intell. Robot. Comput. Vis. XXV Algorithms, Tech. Act. Vis. 6764 (2007) 67640W. https://doi.org/10.1117/12.752808.

[43] T. Jung, Y. Kwon, S. Seo, M.S. Keel, C. Lee, S.H. Choi, S.Y. Kim, S. Cho, Y. Kim, Y.G. Jin, M. Lim, H. Ryu, Y. Kim, J. Kim, C.R. Moon, A 4-tap global shutter pixel with enhanced IR sensitivity for VGA time-of-flight CMOS image sensors, IS T Int. Symp. Electron. Imaging Sci. Technol. 2020 (2020) 1–6. https://doi.org/10.2352/ISSN.2470-1173.2020.7.ISS-103.

[44]　O. Wasenmüller, D. Stricker, Comparison of kinect v1 and v2 depth images in terms of accuracy and precision, Lect. Notes Comput. Sci. (Including Subser. Lect. Notes Artif. Intell. Lect. Notes Bioinformatics). 10117 LNCS (2017) 34–45. https://doi.org/10.1007/978-3-319-54427-4_3.

[45]　Y. Dai, Y. Fu, B. Li, X. Zhang, T. Yu, W. Wang, A new filtering system for using a consumer depth camera at close range, Sensors (Switzerland). 19 (2019) 1–14. https://doi.org/10.3390/s19163460.

[46]　A.P.P. Jongenelen, D.G. Bailey, A.D. Payne, A.A. Dorrington, D.A. Carnegie, Analysis of errors in ToF range imaging with dual-frequency modulation, IEEE Trans. Instrum. Meas. 60 (2011) 1861–1868. https://doi.org/10.1109/TIM.2010.2089190.